\documentclass[12pt]{article}
\begin{document}
\title{Thermal degradation and viscoelasticity
of polypropylene--clay nanocomposites}

\author{A.D. Drozdov\footnote{E--mail:
Aleksey.Drozdov@mail.wvu.edu}\hspace*{1 mm},
A. Al-Mulla, D.A. Drozdov and R.K. Gupta\\
Department of Chemical Engineering\\
West Virginia University\\
P.O. Box 6102\\
Morgantown, WV 26506, USA}
\date{}
\maketitle

\begin{abstract}
Results of torsional oscillation tests are reported
that were performed at the temperature $T=230$~$^{\circ}$C
on melts of a hybrid nanocomposite consisting of
isotactic polypropylene reinforced with 5 wt.\%
of montmorillonite clay.
Prior to mechanical testing, specimens were annealed
at temperatures ranging from $T_{\rm a}=250$
to 310~$^{\circ}$C for various amounts of time
(from 15 to 420 min).
Thermal treatment induced degradation of the matrix
and a pronounced decrease in its molecular weight.
An integro-differential equation is derived for the
evolution of molecular weight based on the
fragmentation--aggregation concept.
This relation involves two adjustable parameters that are
found by fitting observations.
With reference to the theory of transient networks,
constitutive equations are developed for the viscoelastic
response of nanocomposite melts.
The stress--strain relations are characterized by three
material constants (the shear modulus,
the average energy for rearrangement of strands
and the standard deviation of activation energies)
that are determined by matching the dependencies of
storage and loss moduli on frequency of oscillations.
Good agreement is demonstrated between the experimental
data and the results of numerical simulation.
It is revealed that the average energy for separation
of strands from temporary junctions is independent of
molecular weight, whereas the elastic modulus and the
standard deviation of activation energies
linearly increase with mass-average molecular weight.
\end{abstract}
\vspace*{10 mm}

\newpage

\section*{Introduction}

This paper is concerned with (i) the kinetics of thermal
degradation of a hybrid nanocomposite with isotactic
polypropylene (iPP) matrix reinforced with
montmorillonite (MMT) clay and (ii) the effect
of degradation on the viscoelastic response of
the nanocomposite in the melt state.

The choice of iPP for the experimental analysis is explained by
numerous industrial applications of this semicrystalline
polymer (oriented films for packaging, reinforcing fibres,
non-woven fabrics, pipes, etc.).
Montmorillonite is an inorganic clay conventionally used
for preparation of hybrid nanocomposites.
It possesses a layered structure constructed of two
tetrahedral sheets of silica surrounding an octahedral sheet
of alumina or magnesia.
The layers (with a thickness of 1 nm) are stacked by
weak dipole forces, while the galleries between the layers
are occupied by metal cations.

The focus on the influence of thermal degradation
of a polymer--clay nanocomposite on its time-dependent
response may be explained by two reasons:
(i) from the standpoint of applications, the effect
of degradation on the viscoelastic behavior is
of essential importance for the prediction of
mechanical properties of reprocessed industrial
and post-consumer plastic wastes \cite{ISA03};
(ii) from the point of view of fundamental research,
thermal degradation of a polymer is tantamount to
a reduction in its molecular weight \cite{GS85},
which implies that the analysis of the time-dependent
response of nanocomposite melts annealed at
various temperatures $T_{\rm a}$ for various amounts of time
$t_{\rm a}$ sheds some light on their structure--property
relations (correlations between the molecular weight
of the matrix and the material parameters describing
the mechanical response).

The kinetics of thermal and thermo-oxidative degradation of
polypropylene has attracted substantial attention in the
past decade, see \cite{CB97,PVW01,FAV02,GKA03}, to mention a few.
Annealing, thermal degradation and stability of
nanocomposites with a polypropylene matrix and MMT filler
were recently studied in
\cite{MTS01,KKH01,MQH01,GRL01,KKC03,LZW03,HCY03}.
Some relations between the viscoelastic response of iPP--MMT
nanocomposites and their structure have been established
in \cite{KHK97,SAS01,SZW02,EDA03}.

Despite substantial progress in our understanding of
the degradation process and its effect on the mechanical
response of hybrid nanocomposites, it is difficult,
however, to mention a constitutive model that adequately
predicts the time-dependent behavior of iPP--MMT nanocomposite
annealed for a given time $t_{\rm a}$ at a required
temperature $T_{\rm a}$ above the melting temperature
$T_{\rm m}$.

The objective of this study is three-fold:
\begin{enumerate}
\item
To report experimental data in isothermal torsional oscillation
tests with small strains (at the temperature $T=230$~$^{\circ}$C,
which is a typical temperature for injection-molding of iPP)
on specimens annealed for various amounts of time $t_{\rm a}$
(ranging from 15 to 420 min) at various temperatures $T_{\rm a}$
(in the interval from 250 to 310~$^{\circ}$C).

\item
To develop kinetic equations for the evolution of number-average
and mass-average molecular weights at thermal degradation
and to find adjustable parameters in these relations by
fitting the experimental data.

\item
To derive constitutive equations for the viscoelastic
behavior of a nanocomposite melt, to determine
material constants in the stress--strain relations
by matching the dependencies of the storage and loss moduli
on frequency, and to evaluate the effect of thermal
degradation on the time-dependent response.
\end{enumerate}

To make the model tractable from the mathematical
standpoint, we adopt the homogenization hypothesis.
According to it, a complicated micro-structure of
a nanocomposite may be replaced by an equivalent phase,
whose response captures essential features of
the mechanical behavior of the nanocomposite.
As the equivalent phase, a network of macromolecules
is chosen in the present study.

Following \cite{ECH90,HEL91,Dro03}, degradation of the
host matrix in an hybrid nanocomposite is treated as
a combination of two thermally-activated processes:
(i) binary scission of chains,
and (ii) annihilation of end- and side-groups
(diffusion of small-size fragmentation products
and their subsequent evaporation through the surface
of a specimen).
To simplify the analysis, we accept the conventional
assumptions that
(i) the probability of a scission event is independent
of a chain's length and the position of a bond along
the backbone of a chain;
(ii) the diffusivity of separated end- and side-groups
is so large at the exposure temperature $T_{\rm a}$
that the kinetics of diffusion of oligomers
and their evaporation may be disregarded.
As a result, the number of material constants in the
kinetic equation is reduced to two.
These quantities are found by matching a master-curve
for the decrease in the mass-average molecular weight
with exposure time $t_{\rm a}$.

To develop constitutive equations for a nanocomposite
melt, we accept the concept of transient networks
\cite{GT46,Yam56,Lod68,TE92}.
The melt is treated as a network of strands bridged by temporary
junctions (entanglements and physical cross-links
on the surfaces of clay platelets).
At random times, active strands separate from their junctions
being excited by thermal fluctuations, and dangling strands
merge with the network.
Following \cite{DC03a,DC03b}, we assume the network to be strongly
heterogeneous in the sense that different junctions are
characterized by different activation energies for detachment
of strands.
The theory of temporary networks has been applied
to describe the time-depend\-ent behavior of polypropylene
in the solid state in \cite{DC03a,BSC97,SCC99,DC03b},
to mention a few.
The previous studies, however, did not pay much
attention to the effect of molecular weight on
material parameters in the constitutive equations.
The fact that the viscoelastic response of polypropylene
melts is strongly affected by the distribution of
chains' lengths has been established experimentally
in \cite{BMT96,CRG96,SMT01,STM01,FKI02,FI02}.
The aim of this work is to evaluate the influence of
mass-average molecular weight on (i) the concentration
of active strands in an equivalent network,
and (ii) the distribution function for temporary junctions
with various activation energies.

The exposition is organized as follows.
First, experimental data in torsional oscillation tests
are reported on annealed specimens,
and their mass-average molecular weight is
determined by using observations for complex viscosity.
Afterwards, kinetic equations are developed for changes
in the concentration of chains with various lengths
induced by thermal treatment.
Adjustable parameters in these equations are
determined by fitting the observations for molecular
weight.
We proceed with the derivation of stress--strain
relations for an heterogeneous transient network at
three-dimensional deformations with small strains.
The constitutive equations involve three material constants
that are found by matching the data for storage and loss
moduli as functions of frequency of oscillations.
Finally, we establish correlations between the
adjustable parameters in the stress--strain relations
and the mass-average molecular weight of the polypropylene
matrix and discuss the physical meaning of these
relationships.

\section*{Experimental procedure}

Isotactic polypropylene PP 1012 (density 0.906 g/cm$^{3}$,
melt flow rate 1.2 g/10 min) was purchased from
BP Amoco Polymers, Inc.
Concentrate C--44PA containing 43 wt.\% of intercalated
montmorillonite clay in de-agglomerated form
was supplied by Nanocor Co.

To prepare hybrid nanocomposite with 5 wt.\% of MMT clay,
appropriate amounts of isotactic polypropylene and the
concentrate were dried overnight at the temperature
$T=100$~$^{\circ}$C,
pellets were mixed and melt-blended in a twin-screw extruder
(Brabender Instruments, Inc.) with a screw rate of 30 rpm
and temperatures in the extruder barrel of 260, 290, 300
and 290~$^{\circ}$C from hopper to die, respectively.
Strands from the extruder were cooled in a water bath,
cut with a pelletizer,
and dried in an oven at 100~$^{\circ}$C for 12 h.
Circular plates with diameter 64 mm and
thickness 3 mm were molded in injection-molding machine
Battenfeld 1000/315 CDC (Battenfeld).
Specimens for mechanical tests (with diameter 30 mm) were cut
from the plates.

Our choice of the concentration of MMT clay in the hybrid
nanocomposite ($\phi=5$ wt.\%) is explained by the fact
that this amount of nanoclay is sufficient to improve
substantially mechanical properties of neat polypropylene.
According to \cite{SZW02}, the tensile strength of iPP--MMT
nanocomposite reaches its maximum when the clay concentration
equals 5 wt.\%.
Another reason for this choice is that reinforcement of
conventional polymers with MMT clay at higher concentrations
of filler practically does not improve their thermal
stability \cite{PLL03}.

To evaluate the melting temperature $T_{\rm m}$ of the matrix
and the nanocomposite,
DSC (differential scanning calorimetry)
measurements were preformed by using DSC 910S apparatus
(TA Instruments).
The calorimeter was calibrated with indium as a standard.
Two specimens of neat iPP and iPP--MMT nanocomposite
with weights of approximately 13 mg
were tested with a heating rate of 10 K/min from room
temperature to 200~$^{\circ}$C under nitrogen.
The melting temperature $T_{\rm m}=172$~$^{\circ}$C
was determined for isotactic polypropylene as the point
corresponding to the peak on the melting curves.
No substantial changes in the melting temperature were
found for the nanocomposite.
This conclusion is in agreement with the results of
previous studies \cite{HCY03}.

To analyze lattice spacing in the montmorillonite clay
and changes in the crystalline morphology of iPP driven
by the presence of filler,
X--ray diffraction tests were performed on isotactic
polypropylene and polypropylene--clay nanocomposite
by using Rugaku D-max B diffractometer with Cu--K$_{\alpha}$
radiation ($\lambda=1.54$ \AA) generated by a tube with
a voltage of 40 kV and a current of 30 mA.
The Bragg scattering angle ranged from $2\Theta=3$
to $2\Theta=60^{\circ}$ with the step of 0.06$^{\circ}$.
The diffraction spectrum of the hybrid nanocomposite
revealed two peaks: a rather large and wide intensity
maximum at $2\Theta=3.0^{\circ}$
(the basal spacing of 30 \AA) and a smaller one
at $2\Theta=6.5^{\circ}$ (the basal spacing of 17 \AA).
The presence of these peaks indicated a high level of
intercalation of polymer chains between clay platelets.
These observations are similar to those found by other
researchers \cite{KHK97,SZW02,EDA03,ZWF03} on iPP--MMT
nanocomposites with similar concentrations of clay.

Among other interesting features of the X-ray diffractograms,
we would mention:
(i) an increase in $\alpha(110)$ peak of the nanocomposite
compared to that of iPP (by a factor of 5),
(ii) the growth of $\alpha(040)$ peak (by a factor
of 3.3),
(iii) a decrease in $\alpha(130)$ peak (by 26 \%),
(iv) the formation of $\gamma(117)$ peak at $2\Theta=19.6^{\circ}$,
(v) the disappearance of $\alpha(111)$ and $\alpha(-131)$ peaks,
and (vi) a strong growth of $\alpha(060)$ peak
at $2\Theta=25.2^{\circ}$ (by a factor of 3.5).

It is worth noting that a pronounced reduction in the
diffraction peaks at $2\Theta=21.2^{\circ}$ was recently
reported in \cite{YLX03} for hot-stretched polypropylene samples.
It was associated with preferential orientation
of crystallites driven by deformation of specimens.
With reference to this assertion, the fact that
the $\alpha(111)$ and $\alpha(-131)$ peaks were not
observed in the nanocomposite may be ascribed
to the effect of anisotropically distributed clay platelets
on the orientation of lamellar blocks in injection-molded samples.

Dynamic tests were performed by using RMS-800 rheometric
mechanical spectrometer with parallel disks (diameter 25 mm,
gap length 2 mm).
The shear storage modulus $G^{\prime}$ and
the shear loss modulus $G^{\prime\prime}$ were measured
in oscillation tests (the frequency-sweep mode) with
the amplitude of 15 \% and various frequencies $\omega$
ranging from 0.1 to 100 rad/s.
The choice of the amplitude of oscillations was driven
by the following requirements:
(i) mechanical tests were performed in the region of
linear viscoelasticity, and (ii) the torque
was less than its ultimate value 0.2 N$\cdot$m.
The limitation on the minimum frequency of oscillations was
imposed by the condition that the torque exceeded
its minimum value $2.0\cdot 10^{-4}$ N$\cdot$m.
To check that the storage and loss moduli were not affected
by the strain amplitude, several tests were repeated with the
amplitude of 5 \%.
No change in the dynamic moduli was observed.
The temperature in the chamber was controlled with
a standard thermocouple that showed that the temperature
of specimens remained practically constant (with the
accuracy of $\pm 0.5$~$^{\circ}$C).

Prior to mechanical tests, specimens were annealed in
the spectrometer at the temperatures $T_{\rm a}=250$, 270,
290 and 310~$^{\circ}$C (with gap length 3 mm)
for various amounts of time $t_{\rm a}$ ranging from
15 min to 7 h (420 min).
After thermal treatment of a specimen,
the temperature was reduced to the test temperature
$T=230$~$^{\circ}$C,
the specimen was thermally equilibrated (during 5 min),
the gap length was reduced to 2 mm,
an extraneous material was carefully removed,
and the shear storage and loss moduli were measured
at various frequencies $\omega$ starting from the lowest one.
Each test was performed on a new specimen.

We suppose that squeezing of samples between plates of
the spectrometer and removal of the extraneous material
substantially reduced the effect of thermo-oxidative
degradation (compared to that of thermal degradation)
on the mechanical response,
because the major part of the material where
oxidative degradation occurred was taken away
(about one third of the initial mass of each
specimen).
However, we cannot exclude entirely the effect of
diffusion of oxygen to the central part of the samples,
in particular, at the highest temperatures ($T_{\rm a}=290$
and 310~$^{\circ}$C) used in the experiments \cite{HVC94}.

The storage $G^{\prime}$ and loss $G^{\prime\prime}$
moduli, as well as the modulus of the complex viscosity
$\eta$ are plotted versus the logarithm ($\log=\log_{10}$)
of frequency $\omega$ in Figures 1 to 12 (conventional
semi-logarithmic plots are used to characterize changes
in these quantities with frequency).
The shapes of the curves presented coincide qualitatively
with those reported by other researchers, see, e.g.,
\cite{SAS01}.
Given an annealing time $t_{\rm a}$ and an annealing
temperature $T_{\rm a}$, the storage modulus $G^{\prime}$
and the loss modulus $G^{\prime\prime}$ strongly increase
with frequency, whereas the complex viscosity
$\eta$ noticeably decreases with $\omega$.
For a fixed frequency $\omega$, the dynamic moduli
are pronouncedly reduced with $t_{\rm a}$ and $T_{\rm a}$.

\section*{Evaluation of molecular weight}

To assess changes in the molecular weight of the hybrid
nanocomposite induced by thermal degradation of the
matrix, the experimental dependencies $\eta(\omega)$
depicted in Figures 3, 6, 9 and 12 are approximated
by the Cross model
\begin{equation}
\eta(\omega)=\eta_{\infty}+\frac{\Delta \eta}{1
+(\tau \omega)^{\alpha}},
\end{equation}
where $\eta_{\infty}$ is the high-frequency complex viscosity
($\omega\to \infty$),
$\Delta \eta=\eta_{0}-\eta_{\infty}$,
$\eta_{0}$ is the zero-frequency complex viscosity
($\omega=0$),
and $\alpha$ and $\tau$ are adjustable parameters.

Each curve $\eta(\omega)$ is approximated separately.
To find the constants $\eta_{\infty}$, $\Delta \eta$, $\alpha$
and $\tau$ in Eq. (1),
we fix some intervals $[0,\alpha_{\max}]$ and $[0,\tau_{\max}]$,
where the ``best-fit" parameters $\alpha$ and $\tau$
are assumed to be located,
and divide these intervals into $J$ subintervals by
the points $\alpha^{(i)}=i\Delta \alpha$ and
$\tau^{(j)}=j\Delta \tau$ ($i,j=1,\ldots,J-1$) with
$\Delta \alpha=\alpha_{\max}/J$ and $\Delta \tau=\tau_{\max}/J$.
For any pair $\{ \alpha^{(i)}, \tau^{(j)} \}$,
the coefficients $\eta_{\infty}$ and $\Delta\eta$
in Eq. (1) are found by the least-squares method
from the condition of minimum of the function
\[
F=\sum_{\omega_{m}} \Bigl [ \eta_{\rm exp}(\omega_{m})
-\eta_{\rm num}(\omega_{m}) \Bigr ]^{2},
\]
where the sum is calculated over all experimental points
$\omega_{m}$ depicted in Figures 3, 6, 9 and 12,
$\eta_{\rm exp}$ is the complex viscosity
measured in a test,
and $\eta_{\rm num}$ is given by Eq. (1).
The ``best-fit" parameters $\alpha$ and $\tau$ are determined
from the condition of minimum of the function $F$
on the set $ \{ \alpha^{(i)}, \tau^{(j)} \quad (i,j=1,\ldots, J-1)\}$.
After finding the ``best-fit" values $\alpha^{(i)}$ and
$\tau^{(j)}$, this procedure is repeated twice
for the new intervals $[ \alpha^{(i-1)}, \alpha^{(i+1)}]$ and
$[ \tau^{(j-1)}, \tau^{(j+1)}]$,
to ensure an acceptable accuracy of fitting.
Figures 3, 6, 9 and 12 demonstrate good agreement between
the experimental data and the results of numerical simulation.

After finding the zero-frequency viscosity $\eta_{0}$,
the mass-average molecular weight $M_{\rm w}$ is determined
by the conventional equation \cite{Fer80}
\begin{equation}
\frac{\eta_{0}}{\eta_{0}^{\rm ref}}
=\biggl (\frac{M_{\rm w}}{M_{\rm w}^{\rm ref}}\biggr )^{3.4},
\end{equation}
where $\eta_{0}^{\rm ref}$ and $M_{\rm w}^{\rm ref}$
are the zero-frequency viscosity
and the mass-average molecular weight of a reference
(not subjected to thermal treatment) specimen.
The ratio of mass-average molecular weights
\[
d_{\rm w}=\frac{M_{\rm w}}{M_{\rm w}^{\rm ref}}
\]
is found from Eq. (2) by using the experimental data for
$\eta_{0}$.
This method of determining the ratio of mass-average
molecular weights of polypropylene was previously used
in \cite{NW65}.

It is worth noting that Eq. (2) is traditionally employed
for the evaluation of molecular weight of neat polymers.
Its applicability to melts of hybrid nanocomposites is
grounded on the conventional models in rheology of particulate
suspensions \cite{Gup94}, which presume that the relative
viscosity (the ratio of the viscosity of a suspension to
that of the neat polymer melt) is independent of the
melt's structure and is determined by the concentration of
filler exclusively.

The ratio $d_{\rm w}$ is plotted versus annealing time
$t_{\rm a}$ in Figure 13.
With reference to \cite{CB97,Dro03}, we accept the
time--temperature superposition principle for degradation
of the nanocomposite.
According to this hypothesis, the dependencies
$d_{\rm w}(t_{\rm a})$ measured at various annealing
temperatures $T_{\rm a}$ and plotted in semi-logarithmic
coordinates ($d_{\rm w}$ versus $\log t_{\rm a}$)
may be superposed with an acceptable level of accuracy
by shifting the observations along the time-axis.
Applying this approach, we construct the master-curve
depicted in Figure 13.
The experimental data at $T_{\rm a}^{\rm ref}=290$~$^{\circ}$C
are presented without changes.
Observations at the other temperatures
($T_{\rm a}=250$, 270 and 310~$^{\circ}$C)
are shifted along the time-axis by appropriate amounts
$A$ that are determined from the condition that the
experimental data produce a smooth master-curve.

The parameter $A$ is plotted versus the absolute temperature
$T_{\rm a}$ in Figure 14.
The experimental data are approximated by the Arrhenius
dependence
\begin{equation}
\ln A=A_{0}-\frac{A_{1}}{T_{\rm a}},
\end{equation}
where the coefficients $A_{i}$ ($i=0,1$) are determined
by the least-squares method.
Figure 14 demonstrates that Eq. (3) ensures quite acceptable
fit of the observations.

For each annealing time $t_{\rm a}$ and annealing temperature
$T_{\rm a}$, we (i) calculate the material constants $\alpha$
and $\tau$ in Eq. (1) by matching the observations
depicted in Figures 3, 6, 9 and 12,
(ii) find the ratio of mass-average molecular weights
$d_{\rm w}$ from Eqs. (1) and (2),
and (iii) plot the adjustable parameters $\alpha$
and $\tau$ versus $d_{\rm w}$ in Figures 15 and 16.
The experimental data are approximated by the
phenomenological relations
\begin{equation}
\alpha=\alpha_{0}-\alpha_{1}d_{\rm w},
\qquad
\tau=\tau_{0}+\tau_{1}d_{\rm w},
\end{equation}
where the coefficients $\alpha_{i}$ and $\tau_{i}$
($i=0,1$) are found by the least-squares technique.
Figures 15 and 16 reveal that Eq. (4) provides reasonable
quality of matching the observations.
The exponent $\alpha$ in Eq. (1) slightly decreases with
mass-average molecular weight, whereas the characteristic
time $\tau$ strongly grows with $M_{\rm w}$.

\section*{Kinetic equations}

Our aim now is to develop kinetic equations for thermal
degradation of a hybrid nanocomposite and to find adjustable
parameters in these relations by matching the experimental
data depicted in Figure 13.
For this purpose, we (i) adopt a homogenization method,
according to which a complicated micro-structure of a
polymer--clay nanocomposite may be replaced by an equivalent
network of macromolecules, and (ii) accept the
fragmentation--annihilation concept.
The latter means that the degradation process is treated
as a combination of two thermally-induced processes:
binary fragmentation of chains and annihilation
(subsequent breakage, diffusion and evaporation
through the surface of a sample) of end- and side-groups.

\subsection*{Binary scission of chains}

Denote by $\bar{N}(t)$ the number of macromolecules per unit
mass of an equivalent network at an arbitrary instant $t\geq 0$.
Following common practice, we treat chains as sequences
of segments connected by bonds.
Denote by $N_{k}(t)$ is the number of chains (per unit mass)
at time $t$ containing $k$ segments ($k=1,2,\ldots$).
The functions $N_{k}(t)$ obey the conservation law
\begin{equation}
\bar{N}(t)=\sum_{k=1}^{\infty} N_{k}(t).
\end{equation}
Binary scission (fragmentation) of chains is described
by the reactions
\[
N_{k}\to N_{l}+N_{k-l} \qquad (l=1,\ldots, k-1).
\]
Denote by $\gamma$ the rate of scission (the number
of scission events per bond between segments per unit time).
Assuming $\gamma$ to be a function of temperature $T$
only (which implies that $\gamma$ is independent of the
number of segments in a chain),
we arrive at the kinetic equations for the functions $N_{k}(t)$
\begin{equation}
\frac{dN_{k}}{dt}(t) = -\gamma (k-1) N_{k}(t)
+2\gamma \sum_{j=k+1}^{\infty} N_{j}(t) .
\end{equation}
The coefficient $k-1$ in the first term describes
the number of possible scission events in a chain
containing $k$ segments.
The coefficient ``2" before the sum in Eq. (6) indicates
that there are two opportunities (``left" and ``right")
to obtain a chain with $k$ segments after scission of
a chain with a larger number of segments.

\subsection*{Annihilation of chains}

Thermal fluctuations in a network induce not only
binary scission of macromolecules, but also detachment of
end- and side-groups from polymer chains.
As these groups are rather small, they have relatively
large diffusivity, and can easily leave a polymer specimen.
A decrease in a sample's mass with time driven by separation
and subsequent desorption of end- and side-groups and is
treated as their annihilation.

Following \cite{ECH90,HEL91},
we suppose that detachment of small groups within the interval
$[t, t+dt]$ may be thought of as transformation of a chain
with $k$ segments into a chain with $k-1$ segment.
Denote by $\Gamma_{k}$ the ratio of the number of chains
with $k$ segments that lose a segment per unit time
(due to the annihilation process)
to the entire number of these chains $N_{k}$.
The parameter $\Gamma_{k}$ is proportional to the number
of thermal fluctuations (per unit time) that induce
detachment of side-groups and the number of side-groups
per macromolecule.
As both these quantities linearly increase with $k$,
one can write
\begin{equation}
\Gamma_{k}=\Gamma k^{2},
\end{equation}
where $\Gamma$ is a temperature-dependent material parameter.
The kinetic equation for the fragment\-at\-ion--annihilation
process reads
\[
\frac{dN_{k}}{dt}(t)=-\gamma (k-1) N_{k}(t)
+2\gamma \sum_{j=k+1}^{\infty} N_{j}(t)
+\Gamma \Bigl [ (k+1)^{2} N_{k+1}(t)-k^{2} N_{k}(t)\Bigr ] .
\]
Introducing the concentrations of chains with $k$ segments,
\begin{equation}
n_{k}(t)=\frac{N_{k}(t)}{\bar{N}_{0}},
\end{equation}
where $\bar{N}_{0}=\bar{N}(0)$ is the total number of chains
at the initial instant $t=0$, we present this equation
in the form
\begin{equation}
\frac{dn_{k}}{dt}(t) = -\gamma (k-1) n_{k}(t)
+2\gamma \sum_{j=k+1}^{\infty} n_{j}(t)
+\Gamma \Bigl [ (k+1)^{2} n_{k+1}(t)-k^{2} n_{k}(t)\Bigr ] .
\end{equation}

\subsection*{An explicit solution}

Our purpose now is to analyze changes in
the number-average molecular weight $M_{\rm n}$
and the mass-average molecular weight $M_{\rm w}$
determined by the conventional relations
\begin{equation}
M_{\rm n}(t)=\frac{\sum_{k=1}^{\infty} k n_{k}(t)}
{\sum_{k=1}^{\infty} n_{k}(t)},
\qquad
M_{\rm w}(t)(t)=\frac{\sum_{k=1}^{\infty} k^{2} n_{k}(t)}
{\sum_{k=1}^{\infty} k n_{k}(t)},
\end{equation}
when the functions $n_{k}(t)$ are governed by Eq. (9)
with an arbitrary initial condition
\[
n_{k}(0)=n_{0\;k}
\qquad
(k=1,2,\ldots).
\]
Following common practice, it is convenient to suppose
that the number of segments in a chain is large compared
to unity and to replace the discrete index $k$ in Eq. (9)
by a continuous argument $x$.
This results in the integro-differential
equation for the function $n(t,x)$,
\begin{equation}
\frac{\partial n}{\partial t}(t,x) = -\gamma x n(t,x)
+2\gamma \int_{x}^{\infty} n(t,y) dy
+\Gamma \frac{\partial}{\partial x}\Bigl (x^{2} n(t,x) \Bigr ),
\qquad
n(0,x)=n_{0}(x),
\end{equation}
where $n_{0}(x)$ is a given function.
We do not formulate boundary conditions for the function $n(t,x)$,
but assume that this function does not grow very strongly at $x=0$
and decays rapidly at $x\to \infty$ in the sense that the integrals
exist
\begin{equation}
M_{m}(t)=\int_{0}^{\infty} x^{m} n(t,x) dx
\qquad
(m=0,1,2,\ldots).
\end{equation}
Multiplying Eq. (11) by $x^{m}$ ($m=0,1,\ldots$),
integrating over $x$, and using notation (12), we find that
\begin{equation}
\frac{dM_{m}}{dt}(t) = -\gamma M_{m+1}(t)
+2\gamma \int_{0}^{\infty} x^{m} dx \int_{x}^{\infty} n(t,y) dy
+\Gamma \int_{0}^{\infty} x^{m} \frac{\partial}{\partial x}
\Bigl (x^{2} n(t,x) \Bigr ) dx.
\end{equation}
The first integral is transformed by changing the order
of integration,
\[
\int_{0}^{\infty} x^{m} dx \int_{x}^{\infty} n(t,y) dy
=\int_{0}^{\infty} n(t,y)dy \int_{0}^{y} x^{m} dx
=\frac{1}{m+1} \int_{0}^{\infty} y^{m+1} n(t,y) dy
=\frac{1}{m+1} M_{m+1}(t) .
\]
The other integral is calculated by integration by parts,
\[
\int_{0}^{\infty} x^{m} \frac{\partial}{\partial x}
\Bigl (x^{2} n(t,x) \Bigr ) dx
=-m \int_{0}^{\infty} x^{m+1}n(t,x) dx=-mM_{m+1}(t).
\]
Substitution of these expressions into Eq. (13) results in
the differential equation
\begin{equation}
\frac{dM_{m}}{dt}(t)=-\gamma \biggl (\frac{m-1}{m+1}
+ \kappa m \biggr )M_{m+1}(t)
\end{equation}
with
\begin{equation}
\kappa =\frac{\Gamma}{\gamma}.
\end{equation}
Assuming the equivalent network of chains to be monodisperse
at the initial instant $t=0$,
\begin{equation}
n_{0}(x)=\delta (x-L),
\end{equation}
where $L$ is the initial length of chains,
we solve Eq. (14) by Charlesby's method \cite{Cha54}.
As Eq. (14) is linear with respect to the unknown function
$n(t,x)$, appropriate formulas for the moments $M_{m}(t)$
corresponding to an arbitrary initial condition $n_{0}(x)$
are developed by the superposition method.

The $m$th moment $M_{m}(t)$ is expanded into the Taylor
series in time,
\begin{equation}
M_{m}(t)=\sum_{k=0}^{\infty} \frac{M_{m}^{(k)}(0)}{k!}t^{k},
\end{equation}
where $M_{m}^{(k)}(0)$ stands for the $k$th derivative
at the point $t=0$.
Substitution of Eq. (17) into Eq. (14) implies that
\[
M_{m}^{(k)}(0) = (-\gamma)^{k} \prod_{j=0}^{k-1}
\biggl [ \frac{m+j-1}{m+j+1} +\kappa (m+j) \biggr ]
M_{m+k}(0).
\]
It follows from Eqs. (12) and (16) that
\[
M_{m}(0)=L^{m}.
\]
Substitution of these expressions into Eq. (17)
results in
\begin{equation}
M_{m}(t)=L^{m}\biggl [ 1+\sum_{k=1}^{\infty} A_{mk}
(-\gamma t L)^{k} \biggr ],
\end{equation}
where
\[
A_{mk}=\frac{1}{k!}\prod_{j=0}^{k-1}
\biggl [\frac{m+j-1}{m+j+1} + \kappa (m+j) \biggr ].
\]
Introducing the new variable $j^{\prime}=j+1$ and omitting the prime,
we obtain
\begin{equation}
A_{mk} =\prod_{j=1}^{k} \frac{1}{j} \biggl [ \frac{m+j-2}{m+j}
+\kappa (m+j-1) \biggr ].
\end{equation}
Equation (18) implies that for an arbitrary initial
condition $n_{0}(x)$, the moments $M_{m}(t)$ are given by
\begin{equation}
M_{m}(t) = \int_{0}^{\infty} n_{0}(x) x^{m}
\biggl [ 1+\sum_{k=1}^{\infty} A_{mk} (-\gamma t x)^{k}
\biggr ] dx
= M_{m}(0) + \sum_{k=1}^{\infty} A_{mk} M_{m+k}(0)
(-\gamma t )^{k}.
\end{equation}
Although Eqs. (19) and (20) provide explicit expressions
for the moments $M_{m}(t)$, they are not convenient for
the numerical analysis, because the series converges slowly.
These formulas are helpful, however, for the evaluation of
changes in $M_{m}(t)$ at small times, $\gamma t\ll 1$.
Neglecting terms beyond the first order of smallness in Eq. (20)
and using Eqs. (15) and (19), we find that
\begin{eqnarray*}
&& M_{0}(t) = M_{0}(0)+M_{1}(0) \gamma t,
\qquad
M_{1}(t) = M_{1}(0)-M_{2}(0) \Gamma t,
\nonumber\\
&& M_{2}(t) = M_{2}(0)-M_{3}(0) \Bigl (\frac{\gamma}{3}
+2\Gamma \Bigr )t.
\end{eqnarray*}
According to these equations, changes in the moments $M_{0}(t)$
and $M_{1}(t)$ are governed by two different processes:
an increase in $M_{0}(t)$ is driven by fragmentation of
chains, whereas a decrease in $M_{1}(t)$ is induced by
annihilation of end- and side-groups.
Introducing the notation
\begin{equation}
d_{\rm n}(t)=\frac{M_{\rm n}(t)}{M_{\rm n}(0)},
\qquad
d_{\rm w}(t)=\frac{M_{\rm w}(t)}{M_{\rm w}(0)}
\end{equation}
and using Eqs. (10) and (12), we obtain
\begin{eqnarray}
d_{\rm n}(t)=1-\Bigl (M_{\rm n}(0)\gamma+M_{\rm w}(0)\Gamma \Bigr ) t,
\qquad
d_{\rm w}(t)=1-\biggl [ M_{\rm w}(0)\Gamma +M_{\rm z}(0)
\Bigl (\frac{\gamma}{3}+2\Gamma \Bigr )\biggr ] t,
\end{eqnarray}
where $M_{\rm z}(t)=M_{3}(t)/M_{2}(t)$.

\subsection*{Fitting of observations}

As the series in Eq. (18) converges slowly,
we analyze the evolution of the mass-average molecular
weight $M_{\rm w}$ with elapsed time $t$ numerically.
For this purpose, we integrate Eq. (9) with the
initial condition
\begin{equation}
n_{0\;k}=\delta_{kK},
\end{equation}
where $\delta_{ij}$ denotes the Kronecker delta.
Equation (23) corresponds to a monodisperse distribution
of chains in an equivalent network that contain $K$
segments at the initial instant.
We chose this assumption, because the precise initial
distribution of chains in a nanocomposite is unknown.
Bearing in mind that if the maximal number of segments
in a chain equals $K$ at $t=0$,
no chains with higher number of segments can appear at $t>0$
due to the fragmentation--annihilation process and using Eq. (15),
we re-write Eq. (9) as follows:
\begin{equation}
\frac{dn_{k}}{dt}(t) = - \gamma (k-1) n_{k}(t)
+2 \gamma \sum_{j=k+1}^{K} n_{j}(t)
+\kappa \Bigl [ (k+1)^{2} n_{k+1}(t)-k^{2} n_{k}(t)\Bigr ]
\qquad
(k=1,2,\ldots,K).
\end{equation}
We fix the value $K=100$ and integrate Eq. (24)
with the step $\Delta t=0.1$.
This, relatively large, step is chosen because the
fragmentation rate under consideration $\gamma$ is quite small
(of the order of 10$^{-8}$).

To find the adjustable parameters $\gamma$ and $\kappa$,
we fix some intervals $[0,\gamma_{\max}]$ and $[0,\kappa_{\max}]$,
where the ``best-fit" parameters $\gamma$ and
$\kappa$ are assumed to be located,
and divide these intervals into $J$ subintervals by
the points $\gamma^{(i)}=i \Delta \gamma$,
and $\kappa^{(j)}=j \Delta \kappa$ ($i,j=1,\ldots,J-1$)
with $\Delta \gamma=\gamma_{\max}/J$ and
$\Delta \kappa=\kappa_{\max}/J$.
For any pair $\{ \gamma^{(i)}, \kappa^{(j)} \}$,
Eq. (24) with initial condition (23) is integrated
by the Runge--Kutta method.
The best-fit parameters $\gamma$ and $\kappa$ are determined
from the condition of minimum of the function
\[
F=\sum_{t_{m}} \Bigl [ d_{\rm w}^{\;\rm exp}(t_{m})
-d_{\rm w}^{\;\rm num}(t_{m}) \Bigr ]^{2},
\]
where the sum is calculated over all times $t_{m}$
at which observations are presented in Figure 13,
$d_{\rm w}^{\;\rm exp}$ is the ratio of mass-average
molecular weights measured in the tests,
and $d_{\rm w}^{\;\rm num}$ is given by Eqs. (10) and (21).
Figure 13 demonstrates fair agreement between the
observations on specimens annealed at various temperatures
$T_{\rm a}$ and the results of numerical simulation with
$\gamma=7.9\cdot 10^{-8}$ and $\kappa=436.0$.

To ensure the accuracy of numerical simulation, we use three
tests.
First, we verify that at $\kappa=0$, the first moment $M_{1}$
remains independent of time [this conclusion follows from
Eq. (14) with $\kappa=0$].
Secondly, we increase $K$ by twice, decrease the rate of
fragmentation $\gamma$ by twice and check that the
moments $M_{\rm n}(t)$ and $M_{\rm w}(t)$ remain unchanged.
The latter implies that the results of numerical
analysis are independent of our choice of $K=100$.
Finally, we perform simulation at relatively small times
and confirm that the numerical results for the moments
$M_{m}(t)$ ($m=0,1,2$) coincide with analytical solution (22).

The rate of fragmentation $\gamma$ found by matching
observations on iPP--MMT nanocomposite is of the same
order of magnitude as that determined in \cite{Dro03}
for degradation of polystyrene at $T_{\rm a}=275$~$^{\circ}$C.
The parameter $\kappa$ is of the order of 10$^{2}$,
which implies that the influence of annihilation of
side-groups on the degradation process is substantial
\cite{Dro03}.

To compare our results of numerical analysis with
observations reported by other researchers, we recall that
Eq. (3) is based on two hypotheses: (i) the rate of
fragmentation $\gamma$ follows the Arrhenius dependence
on the annealing temperature $T_{\rm a}$,
\begin{equation}
\gamma=\gamma_{0}\exp \Bigl (-\frac{E}{RT_{\rm a}}\Bigr ),
\end{equation}
where $E$ is the activation energy
and $R$ is the universal gas constant,
and (ii) the ratio $\kappa$ of the rates of annihilation
and fragmentation is independent of annealing temperature
$T_{\rm a}$.
It follows from Eq. (25) that the shift factor $A$
\[
A=\frac{\gamma_{0}}{\gamma_{0}^{\rm ref}},
\]
is given by Eq. (3) with
\begin{equation}
\qquad
A_{0}=\frac{E}{R T^{\rm ref}},
\qquad
A_{1}=\frac{E}{R}.
\end{equation}
Calculating the activation energy $E$ from Eq. (26)
and Figure 14, we find that $E=61.8$ kJ/mol.
This value is rather close to the activation energies
for thermal degradation of isotactic polypropylene
$E=50$ to 120 kJ/mol determined in \cite{CB97} based
on results of thermo-gravimetrical tests at low conversion
factors.

\section*{Constitutive equations}

Our aim now is to fit the experimental data for storage
and loss moduli of the hybrid nanocomposite
annealed at various temperatures $T_{\rm a}$.
For this purpose, we derive constitutive equations for the
viscoelastic response of a nanocomposite melt
at three-dimensional deformations with small strains,
simplify these equations for steady shear oscillations,
find adjustable parameters in the stress--strain
relations by matching the observations
depicted in Figures 1, 2, 4, 5, 7, 8, 10 and 11,
and analyze the effect of mass-average molecular weight
on the material constants.
Our analysis is based on the assumption that the
characteristic time for thermal degradation (of
the order of a few hours) substantially exceeds
the characteristic time for relaxation of stresses
in a nanocomposite melt (of the order of a few
seconds), which implies that scission of macromolecules
and annihilation of side-groups may be disregarded
in the analysis of mechanical tests.

With reference to the concept of transient networks,
a nanocomposite melt is treated as an equivalent network of
strands bridged by temporary junctions (entanglements and
physical cross-links whose life-time does not exceed the
characteristic time of a mechanical test).
A strand whose ends are linked to contiguous junctions is
treated as an active one.
When an end of an active strand separates from a junction,
the strand is transformed into the dangling state.
When a free end of a dangling strand captures a nearby
junction, the strand returns into the active state.
Separation of active strands from their junctions and
merging of dangling strands with the network occur at
random times when the strands are excited by thermal
fluctuations.
According to the theory of thermally-activated processes
\cite{Eyr36}, the rate of detachment of strands from
temporary junctions $\Phi$ is governed by the equation
\begin{equation}
\Phi=\Phi_{0} \exp \Bigl (-\frac{\bar{v}}{k_{\rm B}T}\Bigr ),
\end{equation}
where $\Phi_{0}$ is the attempt rate (the number of separation
events per strand per unit time),
$k_{\rm B}$ is Boltzmann's constant,
$T$ is the absolute temperature,
and $\bar{v}\geq 0$ is the activation energy for separation
of an active strand.
The coefficient $\Phi_{0}$ in Eq. (27) is independent of the
activation energy $\bar{v}$ and is determined by
the current temperature $T$ only.
Confining ourselves to isothermal processes at a reference
temperature $T^{\rm ref}$ and introducing the dimensionless activation
energy $v=\bar{v}/(k_{\rm B}T^{\rm ref})$,
we find from Eq. (27) that
\begin{equation}
\Phi(v)=\Phi_{0}\exp (-v).
\end{equation}
To describe the time-dependent response of a nanocomposite melt,
we follow the approach proposed in \cite{DC03a,DC03b}
and suppose that different junctions are characterized by
different dimensionless activation energies $v$.
The distribution of active strands in a transient network
is determined by the number of active strands per unit mass
$\bar{N}_{\rm a}$ and the distribution function $p(v)$.
The quantity $\bar{N}_{\rm a}p(v)dv$ equals the number
of active strands per unit mass linked to junctions with
the dimensionless activation energies $u$ belonging
to the interval $[v,v+dv]$.

Separation of active strands from temporary junctions
and merging of dangling strands with the network are entirely
described by the function $\nu(t,\tau,v)$
that equals the number (per unit mass) of active strands
at time $t\geq 0$ linked to temporary junctions with
activation energy $v$ which have last merged with the network
before instant $\tau\in [0,t]$.

The quantity $\nu(t,t,v)$ equals the number of active strands
(per unit mass) with the activation energy $v$ at time $t$,
\begin{equation}
\nu(t,t,v)=\bar{N}_{\rm a}p(v).
\end{equation}
The function
\begin{equation}
\varphi(\tau,v)=\frac{\partial \nu}{\partial \tau}(t,\tau,v) \biggl
|_{t=\tau}
\end{equation}
determines the rate of reformation for dangling chains:
the amount $\varphi(\tau,v)d\tau$ equals the number of dangling
strands (per unit mass) that merge with temporary junctions
with activation energy $v$ within the interval $[\tau,\tau+d\tau]$.
The quantity
\[
\frac{\partial \nu}{\partial \tau} (t,\tau,v)\;d\tau
\]
is the number of these strands that have not separated from
their junctions during the interval $[\tau, t]$.
The amount
\[
-\frac{\partial \nu}{\partial t} (t,0,v)\;dt
\]
is the number of active strands (per unit mass) that detach (for
the first time) from the network within the interval $[t,t+dt ]$,
while the quantity
\[
-\frac{\partial^{2} \nu}{\partial t\partial \tau} (t,\tau,v)\;dtd\tau
\]
equals the number of strands (per unit mass) that have last merged
with the network within the interval $[\tau,\tau+d\tau ]$ and
separate from the network (for the first time after merging)
during the interval $[t,t+dt ]$.

The rate of detachment $\Phi$ is defined as the ratio of
the number of active strands that separate from temporary
junctions per unit time to the total number of active strands.
Applying this definition to active strands that were connected
with the network at the initial instant $t=0$,
and to those that merged with the network within the interval
$[\tau,\tau +d\tau ]$, we arrive at the differential equations
\begin{equation}
\frac{\partial \nu}{\partial t}(t,0,v)
= - \Phi(v) \nu(t,0,v),
\qquad
\frac{\partial^{2} \nu}{\partial t\partial \tau}(t,\tau,v)
= - \Phi(v) \frac{\partial \nu}{\partial \tau}(t,\tau,v).
\end{equation}
Integration of Eq. (31) with initial conditions (29) (where we set
$t=0$) and (30) implies that
\begin{equation}
\nu(t,0,v) = \bar{N}_{\rm a} p(v) \exp \Bigl [-\Phi(v)t \Bigr ],
\qquad
\frac{\partial \nu}{\partial \tau}(t,\tau,v)
= \varphi(\tau,v) \exp \Bigl [-\Phi(v)(t-\tau) \Bigr ].
\end{equation}
To exclude the function $\varphi(t,v)$ from Eq. (32), we use the
identity
\begin{equation}
\nu(t,t,v)=\nu(t,0,v)+\int_{0}^{t}
\frac{\partial \nu}{\partial \tau}(t,\tau,v) d\tau.
\end{equation}
Substitution of expressions (29) and (32) into Eq. (33) results in
\begin{equation}
\bar{N}_{\rm a} p(v) = \bar{N}_{\rm a} p(v)
\exp \Bigl [-\Phi(v)t \Bigr ]
+\int_{0}^{t} \varphi(\tau,v)\exp \Bigl [-\Phi(v)(t-\tau)\Bigr ]
d\tau.
\end{equation}
The solution of linear integral equation (34) reads
$\varphi(t,v)=\bar{N}_{\rm a} p(v)\Phi(v)$.
It follows from this equality and Eq. (32) that
\begin{equation}
\frac{\partial \nu}{\partial \tau}(t,\tau,v)
= \bar{N}_{\rm a} p(v) \Phi(v) \exp \Bigl [-\Phi(v)(t-\tau) \Bigr ].
\end{equation}

We adopt the conventional assumptions that
(i) the excluded-volume effect and other multi-chain effects
are screened for individual strands by surrounding
macromolecules,
(ii) the energy of interaction between strands can be taken
into account with the help of the incompressibility
condition,
and (iii) thermal oscillations of junctions can be disregarded,
and the strain tensor for the motion of junctions at
the micro-level coincides with the strain tensor for
macro-deformation.

At isothermal deformation with small strains, a strand
is treated as an isotropic incompressible medium.
The strain energy of an active strand $w_{0}$
is determined by the conventional formula
\[
w_{0}=\mu \hat{e}^{\prime}:\hat{e}^{\prime},
\]
where $\mu$ is an average elastic modulus of a strand,
$\hat{e}$ is the strain tensor for transition from the reference
(stress-free) state of the strand to its deformed state,
the prime stands for the deviatoric component of a tensor,
and the colon denotes convolution of two tensors.

According to the affinity hypothesis, the strain energy
$\bar{w}_{0}(t,0)$ of an active strand that has not separated
from the network during the interval $[0,t]$ reads
\[
w(t,0)=\mu \hat{\epsilon}^{\prime}(t):\hat{\epsilon}^{\prime}(t),
\]
where $\hat{\epsilon}(t)$ is the strain tensor for transition from
the initial (stress-free) state of the network to its deformed
state at time $t$.
With reference to \cite{TE92}, we suppose that
stress in a dangling strand totally relaxes before this strand
captures a new junction.
This implies that the stress-free state of an active strand
that merges with the network at time $\tau\geq 0$ coincides
with the deformed state of the network at that instant.
The mechanical energy of an active strand that has
last merged with the network at time $\tau\in [0,t]$ is given by
\[
w(t,\tau)=\mu \Bigl [ \hat{\epsilon}(t)
-\hat{\epsilon}(\tau)\Bigl ]^{\prime}:
\Bigl [ \hat{\epsilon}(t)-\hat{\epsilon}(\tau)\Bigl ]^{\prime}.
\]
Multiplying the strain energy per strand by the number of
active strands per unit mass and summing the mechanical
energies of active strands linked to temporary junctions with
various activation energies,
we find the strain energy per unit mass of an equivalent network
\begin{equation}
W(t)=\mu \int_{0}^{\infty} \biggl \{ \nu(t,0,v)
\hat{\epsilon}^{\prime}(t):\hat{\epsilon}^{\prime}(t)
+\int_{0}^{t} \frac{\partial \nu}{\partial \tau}(t,\tau,v)
\Bigl [ \hat{\epsilon}(t)-\hat{\epsilon}(\tau)\Bigl ]^{\prime}
:\Bigl [ \hat{\epsilon}(t)-\hat{\epsilon}(\tau)\Bigl ]^{\prime}
d\tau \biggr \}dv.
\end{equation}
Differentiating Eq. (36) with respect to time $t$
and using Eqs. (32), (33) and (35), we arrive at the
formula
\begin{equation}
\frac{dW}{dt}(t)=
\hat{A}(t):\frac{d\hat{\epsilon}^{\prime}}{dt}(t)-B(t),
\end{equation}
where
\begin{eqnarray}
\hat{A}(t) &=& 2\mu \bar{N}_{\rm a} \biggl \{ \hat{\epsilon}(t)
-\int_{0}^{t} \hat{\epsilon}(\tau) d\tau
\int_{0}^{\infty} \Phi(v)\exp \Bigl [-\Phi(v)(t-\tau)\Bigr ]
p(v) dv \biggr \}^{\prime},
\\
B(t) &=& \mu \int_{0}^{\infty} \Phi(v) \biggl \{ \nu(t,0,v)
\hat{\epsilon}^{\prime}(t):\hat{\epsilon}^{\prime}(t)
\nonumber\\
&& +\int_{0}^{t} \frac{\partial \nu}{\partial \tau}(t,\tau,v)
\Bigl [ \hat{\epsilon}(t)-\hat{\epsilon}(\tau)\Bigl ]^{\prime}
:\Bigl [ \hat{\epsilon}(t)-\hat{\epsilon}(\tau)\Bigl ]^{\prime}
d\tau \biggr \} dv \geq 0.
\end{eqnarray}
For isothermal deformation of an incompressible medium,
the Clausius--Duhem inequality reads
\[
Q=-\frac{dW}{dt}+\frac{\hat{\sigma}^{\prime}}{\rho}:
\frac{d\hat{\epsilon}^{\prime}}{dt} \geq 0,
\]
where $\rho$ is density,
$Q$ is internal dissipation per unit mass,
and $\hat{\sigma}$ stands for the stress tensor.
Substitution of Eq. (37) into this equation implies that
\begin{equation}
Q(t)=\frac{1}{\rho} \Bigl [\hat{\sigma}^{\prime}(t)
-\rho \hat{A}(t)\Bigr ]: \frac{d\hat{\epsilon}^{\prime}}{dt}(t)
+B(t)\geq 0.
\end{equation}
As the function $B(t)$ is non-negative, see Eq. (39),
dissipation inequality (40) is satisfied, provided that the
expression in the square brackets vanishes.
This assertion together with Eq. (38) results
in the constitutive equation
\begin{equation}
\hat{\sigma}(t) = -P(t)\hat{I}+2G \biggl \{ \hat{\epsilon}^{\prime}(t)
-\int_{0}^{t} \hat{\epsilon}^{\prime}(\tau) d\tau
\int_{0}^{\infty} \Phi(v)\exp \Bigl [-\Phi(v)(t-\tau)\Bigr ]
p(v) dv \biggr \},
\end{equation}
where $P(t)$ is pressure,
$\hat{I}$ is the unit tensor,
and $G=\rho \mu \bar{N}_{\rm a}$ is an analog of the shear
modulus.
Formula (41) describes the time-dependent response of an
equivalent network at arbitrary three-dimensional deformations
with small strains.
In what follows, we confine ourselves to shear tests with
\[
\hat{\epsilon}(t)=\epsilon(t){\bf e}_{1}{\bf e}_{2},
\]
where $\epsilon(t)$ is the shear strain,
and ${\bf e}_{m}$ ($m=1,2,3$) are unit vectors of a Cartesian
frame.
According to Eq. (41), the shear stress $\sigma(t)$ is given by
\begin{equation}
\sigma(t) = 2G \biggl \{ \epsilon(t)
-\int_{0}^{t} \epsilon(\tau) d\tau
\int_{0}^{\infty} \Phi(v)\exp \Bigl [-\Phi(v)(t-\tau)\Bigr ]
p(v) dv \biggr \}.
\end{equation}
It follows from Eq. (42) that in a shear oscillation test with
\[
\epsilon(t)=\epsilon_{0}\exp (i\omega t),
\]
where $\epsilon_{0}$ and $\omega$ are the amplitude and frequency of
oscillations, and $i=\sqrt{-1}$, the transient complex modulus
\[
\bar{G}^{\ast}(t,\omega)=\frac{\sigma(t)}{2\epsilon(t)}
\]
is determined by the formula
\[
\bar{G}^{\ast}(t,\omega)=G\biggl \{ 1-\int_{0}^{\infty} \Phi(v)
p(v) dv \int_{0}^{t}
\exp \Bigl [-\Bigl (\Phi(v)+i\omega \Bigr )s \Bigr ]ds \biggr \},
\]
where $s=t-\tau$.
This equality implies that the steady-state complex modulus
\[
G^{\ast}(\omega)=\lim_{t\to\infty} \bar{G}^{\ast}(t,\omega)
\]
is given by
\[
G^{\ast}(\omega)=G\int_{0}^{\infty}
\frac{i\omega}{\Phi(v)+i\omega} p(v) dv.
\]
This equality together with Eq. (28) implies that
the steady-state storage $G^{\prime}(\omega)$ and loss
$G^{\prime\prime}(\omega)$ shear moduli read
\begin{eqnarray}
G^{\prime}(\omega) &=& G \int_{0}^{\infty}
\frac{\omega^{2}}{\Phi_{0}^{2}\exp(-2v)+\omega^{2}}p(v) dv,
\nonumber\\
G^{\prime\prime}(\omega) &=& G \int_{0}^{\infty} \frac{\Phi_{0}
\exp(-v)\omega}{\Phi_{0}^{2}\exp(-2v)+\omega^{2}}p(v) dv.
\end{eqnarray}
To fit the experimental data, we adopt the random energy
model \cite{Der80} with the quasi-Gaussian distribution
function $p(v)$,
\begin{equation}
p(v) = p_{0}\exp \biggl [
-\frac{(v-V)^{2}}{2\Sigma^{2}}\biggr ]
\quad (v\geq 0),
\qquad
p(v)=0
\quad
(v<0),
\end{equation}
where $V$ and $\Sigma$ are adjustable parameters
(the apparent average activation energy and
the apparent standard deviation of activation
energies, respectively),
and the constant $p_{0}$ is found from the normalization
condition
\begin{equation}
\int_{0}^{\infty} p(v) d v =1.
\end{equation}
Governing equations (43) and (44) involve four material constants:
(i) the instantaneous shear modulus $G$,
(ii) the attempt rate for rearrangement of strands $\Phi_{0}$,
(iii) an analog of the average activation energy
for rearrangement of strands in a network $V$,
and (iv) an analog of the standard deviation of
activation energies $\Sigma$.

When the dimensionless ratio $\xi=\Sigma/V$ is small compared
to unity (it will be shown later that this condition is satisfied
for our experimental data), the number of adjustable
parameters may be reduced to three.
Assuming that
\begin{equation}
\frac{\Sigma}{V}\ll 1,
\end{equation}
we can employ the first equality in Eq. (44) for an arbitrary
(positive and negative) $v$.
Replacing the lower limit of integration in Eqs. (43) by $-\infty$,
we obtain
\begin{eqnarray}
G^{\prime}(\omega) &=& G p_{0} \int_{-\infty}^{\infty}
\frac{\omega^{2}}{\Phi_{0}^{2}\exp(-2v)+\omega^{2}}
\exp \biggl [ -\frac{(v-V)^{2}}{2\Sigma^{2}}\biggr ] dv,
\nonumber\\
G^{\prime\prime}(\omega) &=& G p_{0} \int_{-\infty}^{\infty}
\frac{\Phi_{0} \exp(-v)\omega}{\Phi_{0}^{2}\exp(-2v)
+\omega^{2}}\exp \biggl [-\frac{(v-V)^{2}}{2\Sigma^{2}}\biggr ]dv,
\end{eqnarray}
where
\[
p_{0}=\frac{1}{2\pi\sqrt{\Sigma}}.
\]
To exclude the attempt rate $\Phi_{0}$ from the consideration,
we introduce the notation
\[
\Phi_{0}=\Phi_{\ast}\exp (v_{0}),
\]
where $\Phi_{\ast}$ is a given value
(in what follows, we set $\Phi_{\ast}=10^{10}$ s$^{-1}$),
and $v_{0}=\ln \Phi_{0}/\Phi_{\ast}$.
Substituting this expression into Eq. (47) and introducing the
new variable $v^{\prime}=v-v_{0}$, we find that
\begin{eqnarray*}
G^{\prime}(\omega) &=& G p_{0} \int_{-\infty}^{\infty}
\frac{\omega^{2}}{\Phi_{\ast}^{2}\exp(-2v^{\prime})+\omega^{2}}
\exp \biggl [ -\frac{(v^{\prime}-V^{\prime})^{2}}{2\Sigma^{2}}\biggr ]
dv^{\prime},
\nonumber\\
G^{\prime\prime}(\omega) &=& G p_{0} \int_{-\infty}^{\infty}
\frac{\Phi_{\ast} \exp(-v^{\prime})\omega}{\Phi_{\ast}^{2}\exp(-2v^{\prime})
+\omega^{2}}\exp \biggl [-\frac{(v^{\prime}-V^{\prime})^{2}}{2\Sigma^{2}}\biggr ]
dv^{\prime} ,
\end{eqnarray*}
where
\[
V^{\prime}=V-v_{0}=V-\ln\frac{\Phi_{0}}{\Phi_{\ast}}.
\]
Omitting primes for the sake of simplicity
and replacing the lower limits of integration by zero,
we return to Eqs. (43), where the unknown attempt rate $\Phi_{0}$
is replaced by $\Phi_{\ast}$.
This implies that each set of observations for the
storage and loss shear moduli, $G^{\prime}(\omega)$
and $G^{\prime\prime}(\omega)$, is entirely determined
by three quantities: $G$, $V$ and $\Sigma$.
For hybrid nanocomposites subjected to thermal treatment,
these parameters are functions of annealing temperature $T_{\rm a}$
and annealing time $t_{\rm a}$.

\section*{Fitting of observations}

To assess the effect of temperature and time of
annealing, we determine the quantities $G$, $V$ and $\Sigma$
by matching the experimental data depicted in Figures 1, 2,
4, 5, 7, 8, 10 and 11.
Each set of observations for $G^{\prime}(\omega)$
and $G^{\prime\prime}(\omega)$ is approximated separately.
We fix some intervals $[0,V_{\max}]$ and $[0,\Sigma_{\max}]$,
where the ``best-fit" parameters $V$ and $\Sigma$
are assumed to be located,
and divide these intervals into $J$ subintervals by
the points $V^{(i)}=i\Delta V$ and
$\Sigma^{(j)}=j\Delta \Sigma$ ($i,j=1,\ldots,J-1$) with
$\Delta V=V_{\max}/J$ and $\Delta \Sigma=\Sigma_{\max}/J$.
For any pair $\{ V^{(i)}, \Sigma^{(j)} \}$,
the coefficient $p_{0}$ in Eq. (44) is calculated from Eq. (45),
where the integral is evaluated numerically by Simpson's
method with 400 points and the step $\Delta v=0.1$.
The integrals in Eq. (43) are calculated by using
the same technique.
The shear modulus $G$ is found by the least-squares method
from the condition of minimum of the function
\[
F=\sum_{\omega_{m}} \biggl \{ \Bigl [
G^{\prime}_{\rm exp}(\omega_{m})
-G^{\prime}_{\rm num}(\omega_{m}) \Bigr ]^{2}
+\Bigl [ G^{\prime\prime}_{\rm exp}(\omega_{m})
-G^{\prime\prime}_{\rm num}(\omega_{m}) \Bigr ]^{2} \biggr \},
\]
where the sum is calculated over all experimental points
$\omega_{m}$,
$G_{\rm exp}^{\prime}$ and $G_{\rm exp}^{\prime\prime}$
are the storage and loss moduli measured in a test,
and $G_{\rm num}^{\prime}$ and $G_{\rm num}^{\prime\prime}$
are given by Eq. (43).
The ``best-fit" parameters $V$ and $\Sigma$ are determined
from the condition of minimum of the function $F$
on the set $ \{ V^{(i)}, \Sigma^{(j)} \quad (i,j=1,\ldots, J-1)\}$.
After finding the ``best-fit" values $V^{(i)}$ and
$\Sigma^{(j)}$, this procedure is repeated twice
for the new intervals $[ V^{(i-1)}, V^{(i+1)}]$ and
$[ \Sigma^{(j-1)}, \Sigma^{(j+1)}]$,
to ensure an acceptable accuracy of fitting.
Figures 1, 2, 4, 5, 7, 8, 10 and 11 demonstrate excellent
agreement between the experimental data and the results
of numerical simulation.

For each annealing time $t_{\rm a}$ and annealing temperature
$T_{\rm a}$, we (i) find the ratio of mass-average
molecular weights $d_{\rm w}$ from Eqs. (1) and (2)
and the observations depicted in Figures 3, 6, 9 and 12,
(ii) calculate the material constants $G$, $V$ and $\Sigma$ by
matching the experimental data reported in
Figures 1, 2, 4, 5, 7, 8, 10 and 11, and (iii)
plot the quantities $G$, $V$ and $\Sigma$ versus $d_{\rm w}$ in
Figures 17 and 18.
The experimental data are approximated by the linear
equations
\begin{equation}
G=G_{0}+G_{1}d_{\rm w},
\qquad
V=V_{0},
\qquad
\Sigma=\Sigma_{0}+\Sigma_{1} d_{\rm w},
\end{equation}
where the coefficients $G_{i}$, $V_{i}$ and $\Sigma_{i}$
($i=0,1$) are calculated by the least-squares method.
Figures 17 and 18 show that Eq. (48) correctly describes
changes in the adjustable parameters with mass-average
molecular weight.
The average activation energy for separation of strands
from temporary junctions $V$ is independent of
molecular weight, whereas the shear modulus
$G$ and the standard deviation of activation energies
$\Sigma$ noticeably grow with $M_{\rm w}$.
It can also be seen from Figure 18 that inequality (46)
is satisfied with a reasonable level of accuracy (the
ratio on left-hand side of Eq. (46) does not exceed 0.2).

\section*{Discussion}

We begin with the analysis of material constants in Eq. (1)
that describes the effect of frequency of oscillations
$\omega$ on the modulus of complex viscosity $\eta$.
According to Figure 15, the exponent $\alpha$ is
practically independent of mass-average molecular
weight.
It follows from this conclusion and Eq. (1) that
the curves $\eta(\omega)$ measured at various annealing
temperatures $T_{\rm a}$ and annealing times $t_{\rm a}$
and plotted in double-logarithmic coordinates
may be superposed (with a high level of accuracy)
by shifts along the horizontal (frequency) and
vertical (viscosity) axes.

Figure 16 demonstrates that the characteristic time
$\tau$ linearly decreases with mass-average molecular
weight.
This implies that the reciprocal quantity $\tau^{-1}$
(with the dimension of rate) linearly grows with
$M_{\rm w}$.
The latter conclusion is in agreement with observations
by Bywater and Black \cite{BB65}, who found a similar
trend for the degradation rate of poly(methyl
methacrylate) and poly($\alpha$-methylstyrene).

Figure 17 shows that the elastic modulus $G$ linearly
increases with mass-average molecular weight.
A linear relation between the shear modulus and the
molecular weight between entanglements $M_{\rm e}$
provides a basis for the statistical theory of rubber
elasticity.
It follows from this dependence and Figure 17 that
the molecular weight between entanglements $M_{\rm e}$
is proportional to the mass-average molecular weight
$M_{\rm w}$.
The latter result appears to be quite natural.

According to Figure 18, the average energy for
detachment of strands from temporary junctions $V$
is independent of molecular weight.
This result seems natural as well.
Indeed, for a melt of a hybrid nanocomposite,
the average activation energy $V$ may be treated
as the energy of thermal fluctuations necessary
for mutual displacement of two chains (or for detachment
of a chain from a stack of clay platelets) to a distance
at which the two chains (or the chain and the filler
particle) weakly affect each other.
It seems plausible to assume that this parameter is
independent of the chains' length (because it reflects
local interactions between their segments).
The latter implies that thermal degradation of a nanocomposite
melt (modelled as scission of chains and annihilation
of end- and side groups) should not affect $V$,
which is confirmed by the experimental data presented
in Figure 18.

Figure 18 reveals that the standard deviation of activation
energies $\Sigma$ linearly increases with mass-average
molecular weight.
To provide an explanation for this observation,
we recall that the standard deviation of activation
energies $\Sigma$ may be thought of as a measure of
heterogeneity of an equivalent network of macromolecules,
see Eq. (44).
Thermal degradation of a polymer matrix results in
homogenization of the network (as the rates of
scission of macromolecules and detachment of side-groups
are proportional to chains' lengths), which is observed
as a reduction in $\Sigma$ with a decrease in $d_{\rm w}$.
It is worth noting that a similar decrease in the
inhomogeneity of an equivalent network driven by thermal
degradation of neat iPP was previously observed
as a reduction in the polydispersity index
with annealing time \cite{CB97}.

Figures 15 to 18 show that the adjustable parameters in
the model are not affected by the history of thermal
pre-treatment, but are determined by the current mass-average
molecular weight $M_{\rm w}$ exclusively.
Some scatter should, however, be mentioned of the
experimental data depicted in these figures.
It may be explained by the fact that at each temperature
$T_{\rm a}$ and each annealing time $t_{\rm a}$,
a new sample was used for testing, whose physical properties
do not exactly coincide with those of other specimens.

\section*{Concluding remarks}

A series of torsional oscillation tests with small strains
have been performed at the temperature $T=230$~$^{\circ}$C
on a hybrid nanocomposite with a polypropylene matrix
reinforced with 5 wt.\% of MMT clay.
Prior to mechanical tests, specimens were annealed
at the temperatures $T_{\rm a}=250$, 270, 290 and
310~$^{\circ}$C for various amounts of time $t_{\rm a}$
ranging from 15 to 420 min.
Thermal treatment induced thermal degradation of
samples observed as a pronounced decrease in their
mass-average molecular weight with exposure time
$t_{\rm a}$.

With reference to the fragmentation--annihilation concept,
a kinetic equation has been developed for the concentration
of chains with various lengths.
This relation involves two adjustable parameters that
are found by matching the experimental data
for the evolution of mass-average molecular weight.
In addition to the numerical analysis, an explicit solution
of the kinetic equation has been derived.

A constitutive model has been developed for the viscoelastic
response of a nanocomposite melt at isothermal
three-dimensional deformations with small strains.
The melt is treated as an equivalent transient network
of strands bridged by temporary junctions.
Its time-dependent behavior is modelled as separation of
active strands from their junctions and attachment of dangling
strands to the network.
The rearrangement events occur at random times,
when appropriate strands are thermally activated.

Stress--strain relations for an equivalent heterogeneous
network of strands (where different junctions have different
activation energies for rearrangement of strands)
have been derived by using the laws of thermodynamics.
The constitutive equations involve three material parameters
that are determined by matching the experimental data
for the storage and loss moduli as functions of frequency
of oscillations $\omega$.
Fair agreement is demonstrated between the observations
and the results of numerical simulation.

The following conclusions are drawn:
\begin{enumerate}
\item
The average activation energy for rearrangement of
strands in a transient network $V$ is practically
independent of molecular weight.

\item
The standard deviation of activation energies $\Sigma$
and the shear modulus $G$ linearly grow with mass-average
molecular weight $M_{\rm w}$.

\item
The values of these parameters are independent of
the history of thermal pre-treatment and are entirely
determined by the current mass-average molecular weights.
This implies that thermal degradation may be used
as a quick-and-dirty test directed to establish
correlations between the viscoelastic properties
and molecular weights of polymers and hybrid nanocomposites.
\end{enumerate}

\subsection*{Acknowledgement}

This work was partially supported by the West Virginia
Research Challenge Grant Program.
The authors are deeply indebted to A. Manivannan for
performing WAXD tests.

\newpage
\subsection*{List of figures}

\begin{description}

\item{
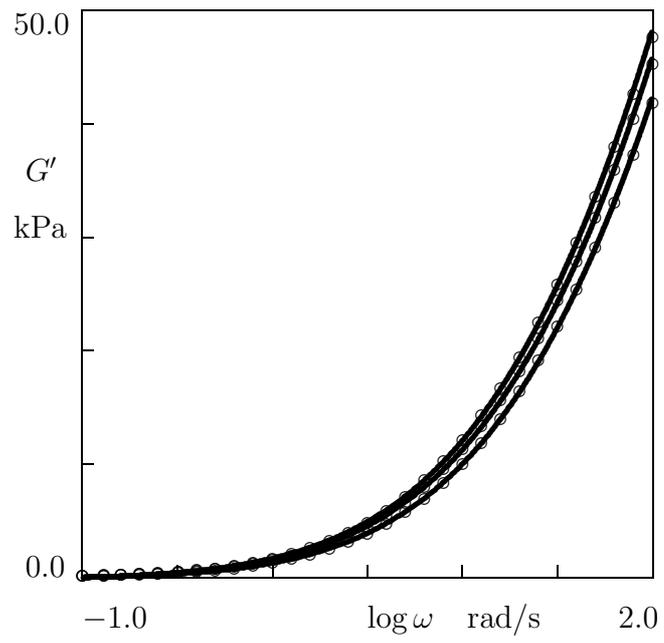
{\bf Figure 1:}
The storage modulus $G^{\prime}$
versus frequency $\omega$.
Circles: experimental data on iPP--MMT nanocomposite
annealed at $T_{\rm a}=250$~$^{\circ}$C for
$t_{\rm a}=0$, 60 and 90 min,
from top to bottom, respectively.
Solid lines: results on numerical simulation}
\vspace*{1 mm}

\item{
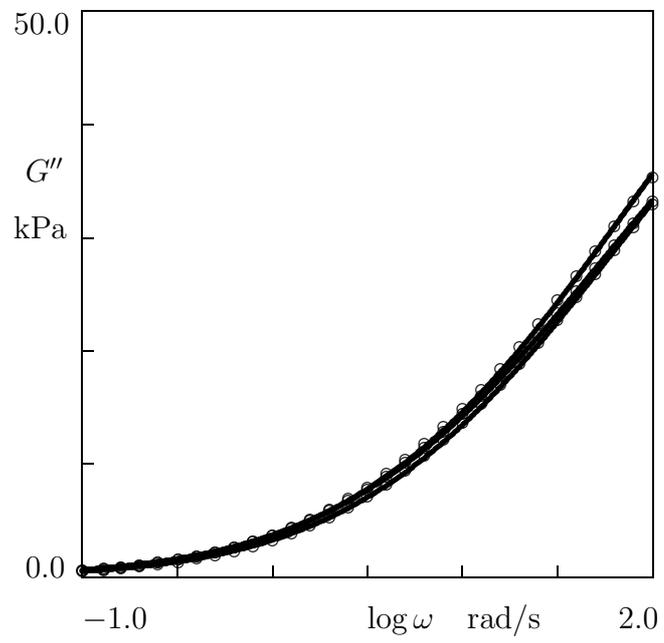
{\bf Figure 2:}
The loss modulus $G^{\prime\prime}$
versus frequency $\omega$.
Circles: experimental data on iPP--MMT
nanocomposite annealed at $T_{\rm a}=250$~$^{\circ}$C
for $t_{\rm a}=0$, 60 and 90 min, from top to bottom,
respectively.
Solid lines: results on numerical simulation}
\vspace*{1 mm}

\item{
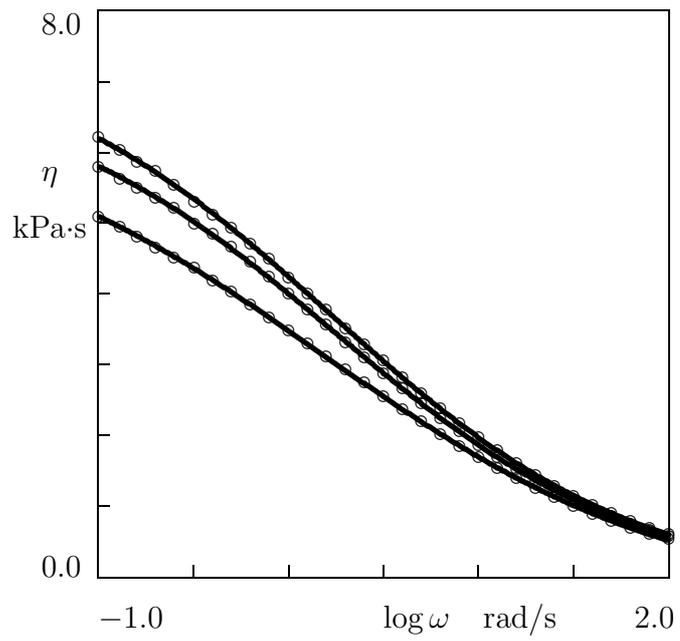
{\bf Figure 3:}
The complex viscosity $\eta$ versus frequency $\omega$.
Circles: experimental data on iPP--MMT nanocomposite
annealed at $T_{\rm a}=250$~$^{\circ}$C
for $t_{\rm a}=0$, 60 and 90 min, from top to bottom,
respectively.
Solid lines: results on numerical simulation}
\vspace*{1 mm}

\item{
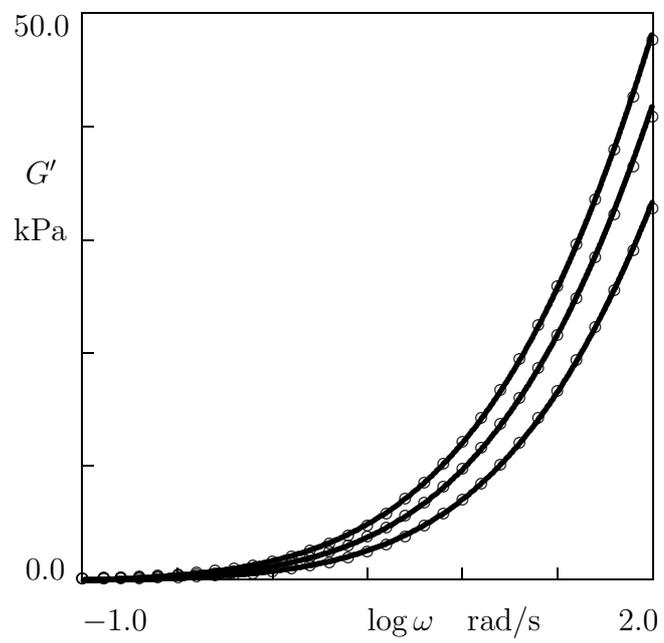
{\bf Figure 4:}
The storage modulus $G^{\prime}$
versus frequency $\omega$.
Circles: experimental data on iPP--MMT nanocomposite
annealed at $T_{\rm a}=270$~$^{\circ}$C
for $t_{\rm a}=0$, 60 and 90 min.
Solid lines: results on numerical simulation}
\vspace*{1 mm}

\item{
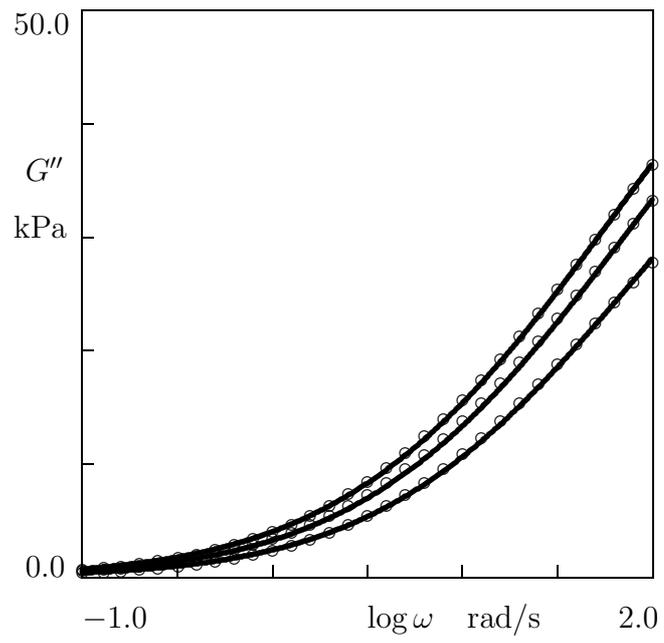
{\bf Figure 5:}
The loss modulus $G^{\prime\prime}$ versus frequency $\omega$.
Circles: experimental data on iPP--MMT nanocomposite
annealed at $T_{\rm a}=270$~$^{\circ}$C
for $t_{\rm a}=0$, 60 and 90 min.
Solid lines: results on numerical simulation}
\vspace*{1 mm}

\item{
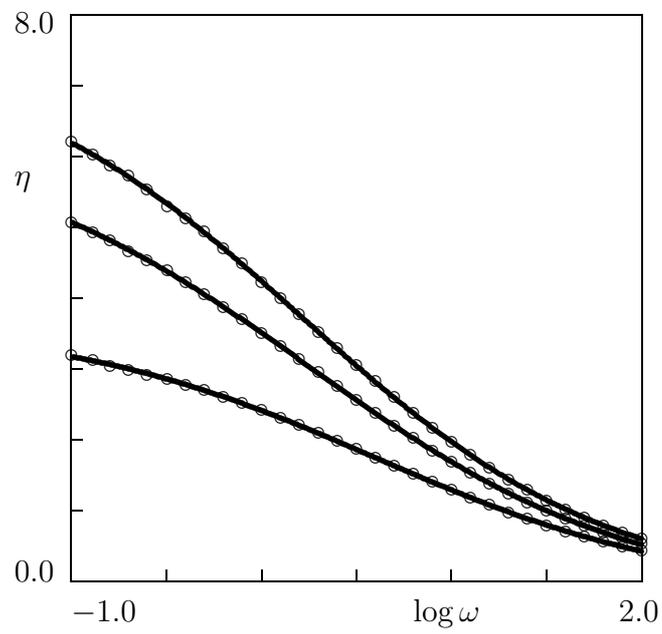
{\bf Figure 6:}
The complex viscosity $\eta$ versus frequency $\omega$.
Circles: experimental data on iPP--MMT annealed at
$T_{\rm a}=270$~$^{\circ}$C for $t_{\rm a}=0$, 60 and
90 min, from top to bottom, respectively.
Solid lines: results on numerical simulation}
\vspace*{1 mm}

\item{
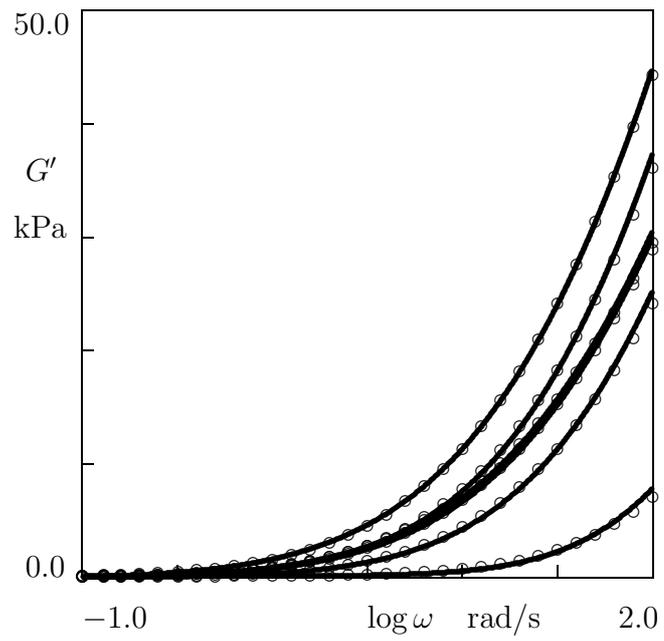
{\bf Figure 7:}
The storage modulus $G^{\prime}$ versus frequency $\omega$.
Circles: experimental data on iPP--MMT nanocomposite
annealed at $T_{\rm a}=290$~$^{\circ}$C
for $t_{\rm a}=0$, 15, 30, 60, 120 and 420 min,
from top to bottom, respectively.
Solid lines: results on numerical simulation}
\vspace*{1 mm}

\item{
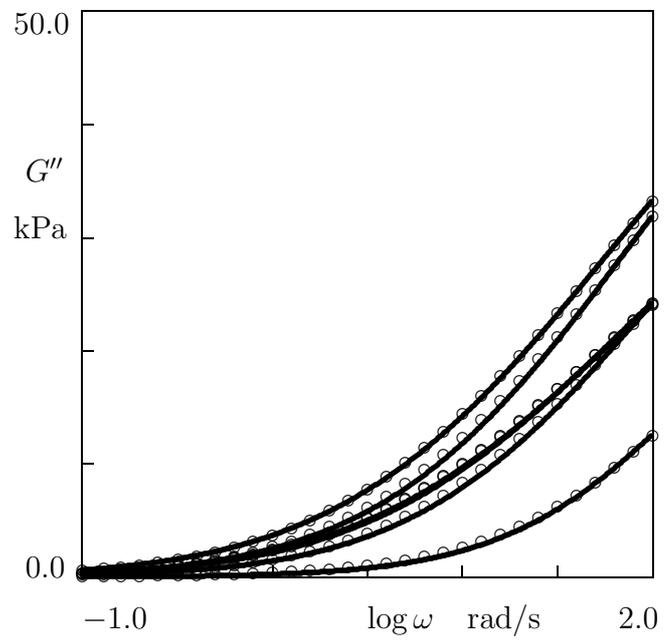
{\bf Figure 8:}
The loss modulus $G^{\prime\prime}$ versus frequency $\omega$.
Circles: experimental data on iPP--MMT nanocomposite
annealed at $T_{\rm a}=290$~$^{\circ}$C
for $t_{\rm a}=0$, 15, 30, 60, 120 and 420 min,
from top to bottom, respectively
Solid lines: results on numerical simulation}
\vspace*{1 mm}

\item{
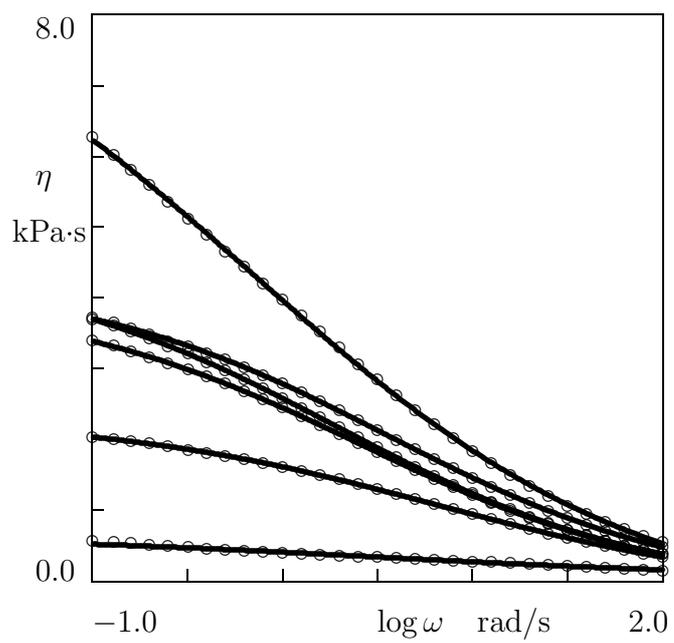
{\bf Figure 9:}
The complex viscosity $\eta$ versus frequency $\omega$.
Circles: experimental data on iPP--MMT nanocomposite
annealed at $T_{\rm a}=290$~$^{\circ}$C
for $t_{\rm a}=0$, 15, 30, 60, 120 and 420 min,
from top to bottom, respectively.
Solid lines: results on numerical simulation}
\vspace*{1 mm}

\item{
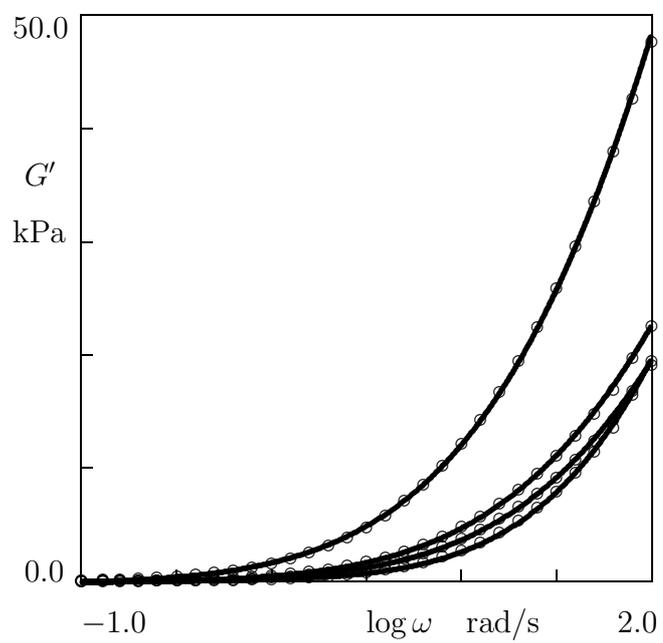
{\bf Figure 10:}
The storage modulus $G^{\prime}$ versus frequency $\omega$.
Circles: experimental data on iPP--MMT nanocomposite
annealed at $T_{\rm a}=310$~$^{\circ}$C for $t_{\rm a}=0$,
15, 30 and 60 min, from top to bottom, respectively.
Solid lines: results on numerical simulation}
\vspace*{1 mm}

\item{
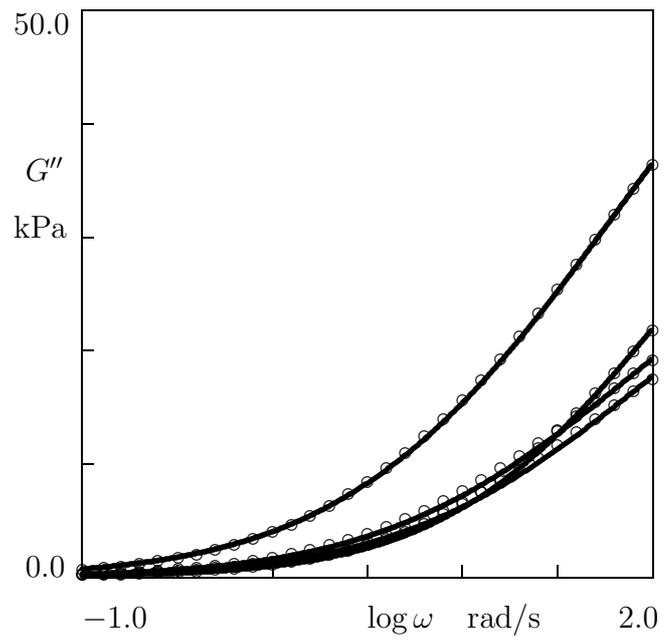
{\bf Figure 11:}
The loss modulus $G^{\prime\prime}$ versus frequency $\omega$.
Circles: experimental data on iPP--MMT nanocomposite
annealed at $T_{\rm a}=310$~$^{\circ}$C for
$t_{\rm a}=0$, 15, 30 and 60 min,
from top to bottom, respectively.
Solid lines: results on numerical simulation}
\vspace*{1 mm}

\item{
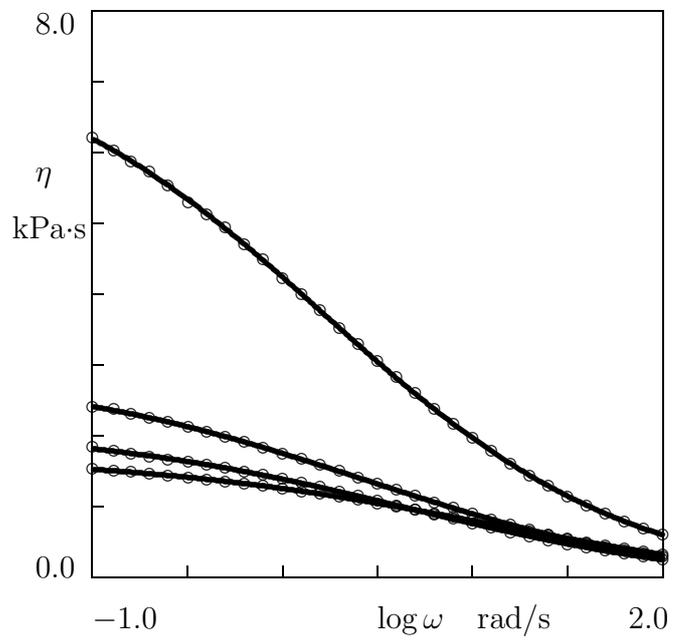
{\bf Figure 12:}
The complex viscosity $\eta$ versus frequency $\omega$.
Circles: experimental data on iPP--MMT nanocomposite
annealed at $T_{\rm a}=310$~$^{\circ}$C
for $t_{\rm a}=0$, 15, 30 and 60 min,
from top to bottom, respectively.
Solid lines: results on numerical simulation}
\vspace*{1 mm}

\item{
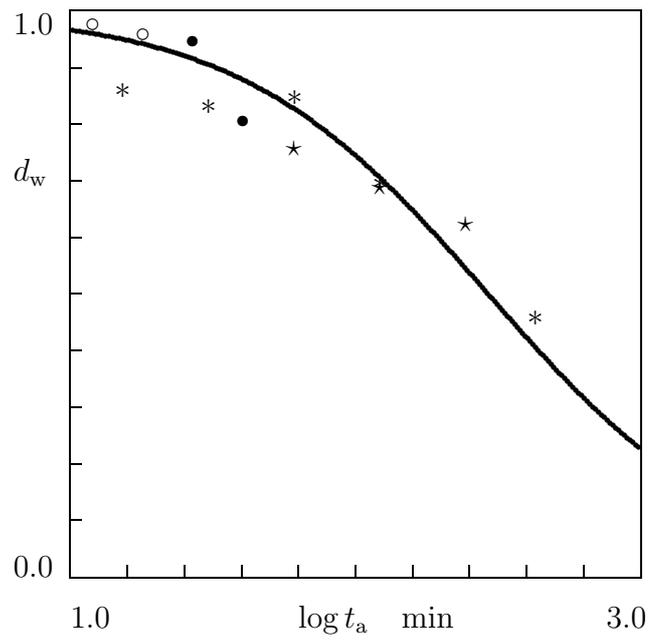
{\bf Figure 13:}
The ratio of mass-average molecular weights
$d_{\rm w}$ versus annealing time $t_{\rm a}$.
Symbols: experimental data on iPP--MMT nanocomposite
annealed at temperatures $T_{\rm a}$~$^{\circ}$C.
Unfilled circles: $T_{\rm a}=250$;
filled circles: $T_{\rm a}=270$;
asterisks: $T_{\rm a}=290$;
stars: $T_{\rm a}=310$.
Solid line: results of numerical simulation}
\vspace*{1 mm}

\item{
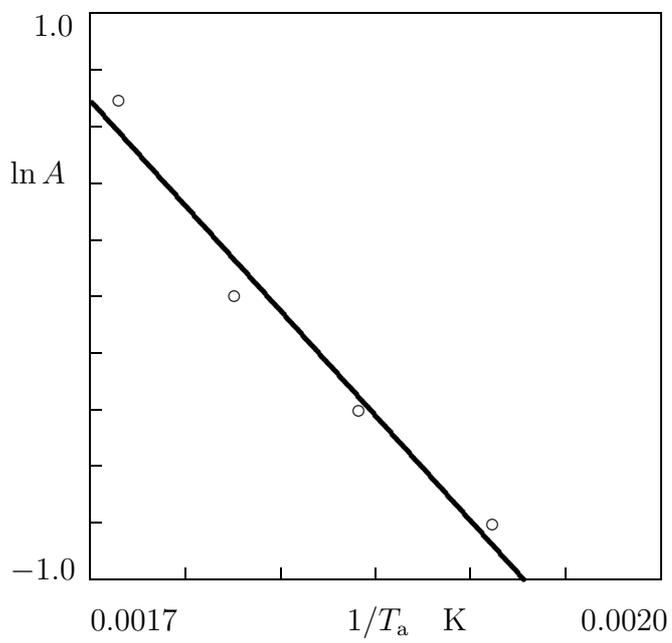
{\bf Figure 14:}
The shift factor $A$ versus annealing temperature $T_{\rm a}$.
Circles: experimental data on iPP--MMT nanocomposite.
Solid line: approximation of the experimental data
by Eq. (3) with $A_{0}=13.32$ and $A_{1}=7.43\cdot 10^{3}$}
\vspace*{1 mm}

\item{
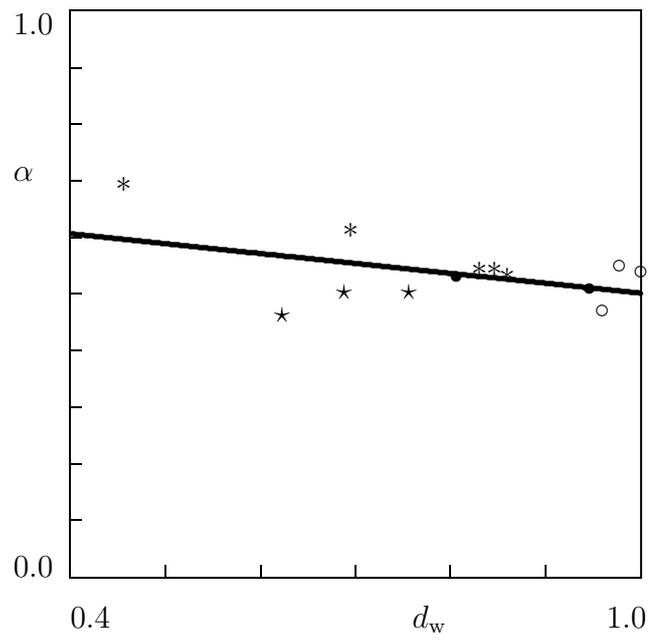
{\bf Figure 15:}
The dimensionless exponent $\alpha$ versus the ratio
of mass-average molecular weights $d_{\rm w}$.
Symbols: treatment of observations
on iPP--MMT nanocomposite at annealing
temperatures $T_{\rm a}$~$^{\circ}$C.
Unfilled circles: $T_{\rm a}=250$;
filled circles: $T_{\rm a}=270$;
asterisks: $T_{\rm a}=290$;
stars: $T_{\rm a}=310$.
Solid line: approximation of the experimental data
by Eq. (4) with $\alpha_{0}=0.68$ and $\alpha_{1}=0.17$}
\vspace*{1 mm}

\item{
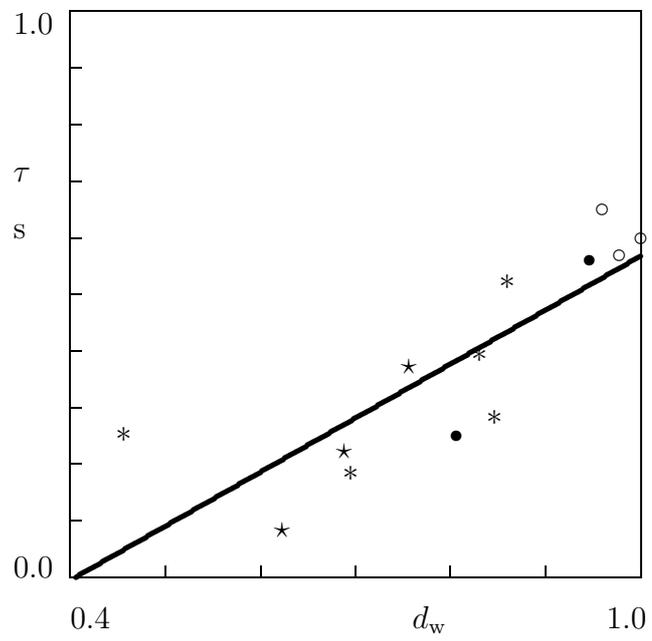
{\bf Figure 16:}
The characteristic time $\tau$ versus the ratio
of mass-average molecular weights $d_{\rm w}$.
Symbols: treatment of observations
on iPP--MMT nanocomposite at annealing temperatures
$T_{\rm a}$~$^{\circ}$C.
Unfilled circles: $T_{\rm a}=250$;
filled circles: $T_{\rm a}=270$;
asterisks: $T_{\rm a}=290$;
stars: $T_{\rm a}=310$.
Solid line: approximation of the experimental data
by Eq. (4) with $\tau_{0}=-0.39$ and $\tau_{1}=0.95$}
\vspace*{1 mm}

\item{
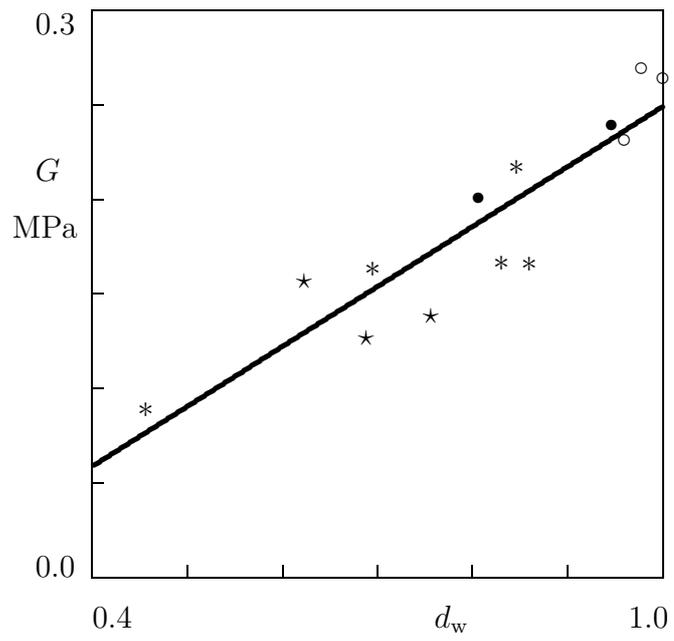
{\bf Figure 17:}
The instantaneous shear modulus $G$ versus the ratio
of mass-average molecular weights $d_{\rm w}$.
Symbols: treatment of observations on iPP--MMT
nanocomposite at annealing temperatures $T_{\rm a}$~$^{\circ}$C.
Unfilled circles: $T_{\rm a}=250$;
filled circles: $T_{\rm a}=270$;
asterisks: $T_{\rm a}=290$;
stars: $T_{\rm a}=310$.
Solid line: approximation of the experimental data
by Eq. (48) with $G_{0}=-0.07$ and $G_{1}=0.32$}
\vspace*{1 mm}

\item{
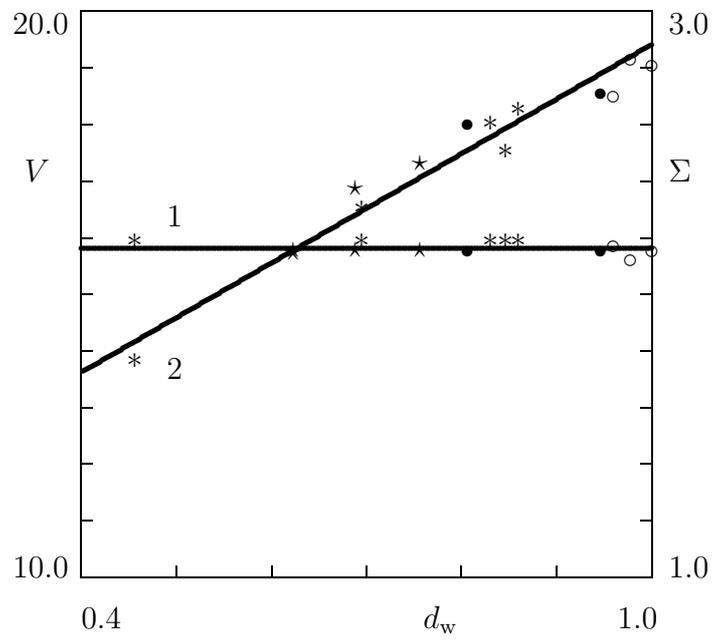
{\bf Figure 18:}
The average activation energy for rearrangement
of strands $V$ and the standard deviation of activation
energies $\Sigma$ versus the ratio of mass-average
molecular weights $d_{\rm w}$.
Symbols: treatment of observations on iPP--MMT
nanocomposite at annealing temperatures
$T_{\rm a}$~$^{\circ}$C.
Unfilled circles: $T_{\rm a}=250$;
filled circles: $T_{\rm a}=270$;
asterisks: $T_{\rm a}=290$;
stars: $T_{\rm a}=310$.
Solid lines: approximation of the experimental data
by Eq. (48) with $V_{0}=15.82$ (curve 1) and
$\Sigma_{0}=0.95$ and $\Sigma_{1}=1.93$ (curve 2)}
\end{description}


\newpage
\begin{thebibliography}{50}

\bibitem{ISA03}
Incarnato, L.; Scarfato, P.; Acierno, D.; Milana, M.R.; Feliciani, R.:
J Appl Polym Sci 2003, 89, 1768--1778.

\bibitem{GS85}
Grassie, N.; Scott, G.:
Polymer Degradation and Stabilisation.
Cambridge Univ. Press, Cambridge 1985.

\bibitem{CB97}
Chan, J.H.; Balke, S.T.:
Polym Degrad Stab 1997, 57, 113--125, 127--134, 135--149.

\bibitem{PVW01}
Peterson, J.D.; Vyazovkin, S.; Wight, C.A.:
Macromol Chem Phys 2001, 202, 775--784.

\bibitem{FAV02}
Fayolle, B.; Audouin, L.; Verdu, J.:
Polym Degrad Stab 2002, 75, 123--129.

\bibitem{GKA03}
Gao, Z.; Kaneko, T.; Amasaki, I.; Nakada, M.:
Polym Degrad Stab 2003, 80, 269--274.

\bibitem{MTS01}
Manias, E.; Touny, A.; Wu, L.; Strawhecker, K.; Lu, B.; Chung, T.C.:
Chem Mater 2001, 13, 3516--3523.

\bibitem{KKH01}
Kodgire, P.; Kalgaonkar, R.; Hambir, S.; Bulakh, N.; Jog, J.P.:
J Appl Polym Sci 2001, 81, 1786-1792.

\bibitem{MQH01}
Ma, J.; Qi, Z.; Hu, Y.:
J Appl Polym Sci 2001, 82, 3611--3617.

\bibitem{GRL01}
Galgali, G.; Ramesh, C.; Lele, A.:
Macromolecules 2001, 34, 852--858.

\bibitem{KKC03}
Koo, C.M.; Kim, M.J.; Choi, M.H.; Kim, S.O.; Chung, I.J.:
J Appl Polym Sci 2003, 88, 1526--1535.

\bibitem{LZW03}
Li, J.; Zhou, C.; Wang, G.; Zhao, D.:
J Appl Polym Sci 2003, 89, 318--323.

\bibitem{HCY03}
He, J.-D.; Cheung, M.K.; Yang, M.-S.; Qi, Z.:
J Appl Polym Sci 2003, 89, 3404-3415.

\bibitem{KHK97}
Kawasumi, M.; Hasegawa, N.; Kato, M.; Usuki, A.; Okada, A.:
Macromolecules 1997, 30, 6333--6338.

\bibitem{SAS01}
Solomon, M.J.; Almussallam, A.S.; Seefeldt, K.F.;
Somwangthanaroj, A.; Varadam, P.:
Macromolecules 2001, 34, 1864--1872.

\bibitem{SZW02}
Svoboda, P.; Zeng, C.; Wang, H.; Lee, L.J.; Tomasko, D.L.:
J Appl Polym Sci 2002, 85, 1562--1570.

\bibitem{EDA03}
Ellis, T.S.; D`Angelo, J.S.:
J Appl Polym Sci 2003, 90, 1639--1647.

\bibitem{ECH90}
Edwards, B.F.; Cai, M.; Han, H.:
Phys Rev A 1990, 41, 5755--5757.

\bibitem{HEL91}
Huang, J.; Edwards, B.F.; Levine, A.D.:
J Phys A: Math Gen 1991, 24, 3967--3977.

\bibitem{Dro03}
Drozdov, A.D.:
cond-mat/0309677.

\bibitem{GT46}
Green, M.S.; Tobolsky, A.V.:
J Chem Phys 1946, 14, 80--92.

\bibitem{Yam56}
Yamamoto, M.:
J Phys Soc Japan 1956, 11, 413--421.

\bibitem{Lod68}
Lodge, A.S.:
Rheol Acta 1968, 7, 379--392.

\bibitem{TE92}
Tanaka, F.; Edwards, S.F.:
Macromolecules 1992, 25, 1516--1523.

\bibitem{DC03a}
Drozdov, A.D.; Christiansen, J. deC.:
Int J Solids Struct 2003, 40, 1337--1367.

\bibitem{DC03b}
Drozdov, A.D.; Christiansen, J. deC.:
Polym Eng Sci 2003, 43, 946--959.

\bibitem{BSC97}
Bensason, S.; Stepanov, E.V.; Chum, S.; Hiltner, A.; Baer, E.:
Macromolecules 1997, 30, 2436--2444.

\bibitem{SCC99}
Sweeney, J.; Collins, T.L.D.; Coates, P.D.; Duckett, R.A.:
J Appl Polym Sci 1999, 72, 563--575.

\bibitem{BMT96}
Barakos, G.; Mitsoulis, E.; Tzoganakis, C.; Kajiwara, T.:
J Appl Polym Sci 1996, 59, 543--556.

\bibitem{CRG96}
Carrot, C.; Revenu, P.; Guillet, J.:
J Appl Polym Sci 1996, 61, 1887--1897.

\bibitem{SMT01}
Sugimoto, M.; Masubuchi, Y.; Takimoto, J.; Koyama, K.:
J Polym Sci Part B: Polym Phys 2001, 39, 2692--2704.

\bibitem{STM01}
Sugimoto, M.; Tanaka, T.; Masubuchi, Y.; Takimoto, J.-I.; Koyama, K.:
J Appl Polym Sci 2001, 73, 1493--1500.

\bibitem{FKI02}
Fujiyama, M.; Kitajima, Y.; Inata, H.:
J Appl Polym Sci 2002, 84, 2128--2141.

\bibitem{FI02}
Fujiyama, M.; Inata, H.:
J Appl Polym Sci 2002, 84, 2157--2170.

\bibitem{PLL03}
Pramoda, K.P.; Liu, T.; Liu, Z.; He, C.; Sue, H.-J.:
Polym Degrad Stab 2003, 81, 37--56.

\bibitem{ZWF03}
Zhang, Q.; Wang, Y.; Fu, Q.:
J Polym Sci Part B: Polym Phys 2003, 41, 1--10.

\bibitem{YLX03}
Yang, W.; Li, Z.-M.; Xie, B.-H.; Feng, J.-M.; Shi, W.; Yang, M.-B.:
J Appl Polym Sci 89, 2003, 686--690.

\bibitem{HVC94}
Horrocks, A.R.; Valinejad, K.; Crighton, J.S.:
J Appl Polym Sci 1994, 54, 593--600.

\bibitem{Fer80}
Ferry, J.D.:
Viscoelastic Properties of Polymers.
Wiley, New York 1980.

\bibitem{NW65}
Nakajima, N.; Wong, P.S.L.:
J Appl Polym Sci 1965, 9, 3141--3152.

\bibitem{Gup94}
Gupta, R.K.:
In Advani, S.G. (Ed.):
Flow and Rheology in Polymer Composites Manifacturing.
Elsevier, Amsterdam 1994, pp. 9--51.

\bibitem{Cha54}
Charlesby, A.:
Proc Roy Soc London A 1954, 224, 120.

\bibitem{Eyr36}
Eyring, H.:
J Chem Phys 1936, 4, 283--291.

\bibitem{Der80}
Derrida, B.
Phys Rev Lett 1980, 45, 79--92.

\bibitem{BB65}
Bywater, S.; Black, P.E.:
J Phys Chem 1965, 69, 2967--2970.

\end{thebibliography}
\end{document}